\documentclass[11pt]{article}
\pdfoutput=1 
\usepackage[super,sort&compress,comma]{natbib} 
\usepackage[version=3]{mhchem}
\usepackage[left=2.0cm, right=2.0cm, top=2.0cm, bottom=3.0cm]{geometry}
\usepackage{times,mathptmx}
\usepackage{sectsty}
\usepackage{graphicx}
\usepackage{float}
\usepackage[english]{babel}
\usepackage{hyperref}
\usepackage{epstopdf}
\usepackage{amssymb}
\usepackage{pifont}
\usepackage{authblk}

\title{Striking light-induced nonadiabatic fingerprint in the low-energy vibronic spectra of polyatomic molecules}
\author[1,2]{Csaba F\'abri}
\author[3]{Benjamin Lasorne}
\author[4]{G\'abor J. Hal\'asz}
\author[5]{Lorenz S. Cederbaum}
\author[6,7]{\'Agnes Vib\'ok}
\affil[1]{Laboratory of Molecular Structure and Dynamics, Institute of Chemistry, E\"otv\"os Lor\'and University, P\'azm\'any P\'eter s\'et\'any 1/A, H-1117 Budapest, Hungary; E-mail: ficsaba@caesar.elte.hu}
\affil[2]{MTA-ELTE Complex Chemical Systems Research Group, P.O. Box 32, H-1518 Budapest 112, Hungary}
\affil[3]{Institut Charles Gerhardt Montpellier (ICGM), Universit\'e de Montpellier, CNRS, ENSCM, F-34095 Montpellier, France}
\affil[4]{Department of Information Technology, University of Debrecen, P.O. Box 400, H-4002 Debrecen, Hungary}
\affil[5]{Theoretische Chemie, Physikalish-Chemisches Institut, Universit\"at Heidelberg, Im Neuenheimer Feld 229, 69120 Heidelberg, Germany}
\affil[6]{Department of Theoretical Physics, University of Debrecen, PO Box 400, H-4002 Debrecen, Hungary; E-mail: vibok@phys.unideb.hu}
\affil[7]{ELI-ALPS, ELI-HU Non-Profit Ltd, H-6720 Szeged, Dugonics t\'er 13, Hungary}
\date{\today}

\begin{document}

\begin{titlepage}
\maketitle
\end{titlepage}

\begin{abstract}
Nonadiabaticity, i.e., the effect of mixing electronic states by
nuclear motion, is a central phenomenon in molecular science. The
strongest nonadiabatic effects arise due to the presence of conical
intersections of electronic energy surfaces. These intersections are
abundant in polyatomic molecules. Laser light can induce in a controlled
manner new conical intersections, called light-induced conical
intersections, which lead to strong nonadiabatic effects similar to
those of the natural conical intersections. These effects are, however,
controllable and may even compete with those of the natural
intersections. In this work we show that the standard low-energy
vibrational spectrum of the electronic ground state can change
dramatically by inducing nonadiabaticity via a light-induced conical
intersection. This generic effect is demonstrated for an explicit
example by full-dimensional high-level quantum calculations using a
pump-probe scheme with a moderate-intensity pump laser and a weak probe
laser.
\end{abstract}

\section{Introduction}
Nonadiabatic phenomena\cite{Lenz1,Yarkony,Baer1,Lenz2,Domcke1,Baer2}
occurring in polyatomic molecules have been known for decades to play a fundamental role
in most photobiological, photochemical, and photophysical
processes.\cite{Kim,Polli,Worner,You,Musser,Conta,Rebeca,08AsDeDi,10Martinez,18CuMa,15KoBeDo,18BeKoRo}
Such effects become significant around molecular geometries where two (or more)
potential energy surfaces (PESs) are energetically degenerate, such that
the Born--Oppenheimer (BO) approximation\cite{Born1} breaks down.
This yields strong mixing among BO electronic states and subsequent radiationless
transitions correlated with variations of the molecular geometry (nuclear degrees of freedom).

Such degeneracy points, termed conical intersections (CIs), are characterized locally
by a two-dimensional subspace within the space spanned by the nuclear degrees of freedom,
the so-called ``branching'' space (BS) along which the degeneracy is lifted to first order around the CI.
This can occur only in molecules with more than two atoms (Wigner's noncrossing rule).\cite{29NeWi}
Indeed, diatomic molecules have a single vibrational coordinate, which cannot accommodate both
zero energy-difference tuning and zero coupling except for symmetry reasons.
CIs provide ultrafast nonradiative decay channels for electronically excited molecules.
In the vicinity of such points, nonadiabatic dynamics (beyond BO) takes place typically
on the femtosecond time scale ($\sim$10--100 fs). While it is by now well established that they are present
in small- or medium-sized molecules, CIs have been shown to be ubiquitous
in large molecular systems, hence playing a key role in most photoinduced processes. 

More than a decade ago, it has been demonstrated that laser light with significant intensity
(dressed-state regime) can in turn induce CI-type situations between field-dressed PESs
and thus give rise to new types of nonadiabatic phenomena,
even in diatomic molecules.\cite{Lenz3,Lenz4}
Such points were then termed light-induced conical intersections (LICIs).
Whereas the location of a ``natural'' CI in a field-free polyatomic molecule
and the strength of the nonadiabatic coupling are inherent properties of the system,
the occurrence of the crossing point is determined by the laser frequency while the strength of
the nonadiabatic-type coupling is controlled by the laser intensity. 

Several theoretical and experimental works have demonstrated on both fronts that LICIs
give rise to a variety of unexpected nonadiabatic phenomena in molecules,\cite{Gabi1,Gabi2,Gabi3,Gabi4,Gabi5}
with significant impact on various dynamical\cite{Gabi6,Gabi7,Gabi8,Book}
and spectroscopic properties.\cite{Gabi9,Gabi10,Gabi11,Gabi12}
These works were mainly focused on diatomic molecules where the angle between the internuclear axis and the field polarization axis
provides the missing degree of freedom for a CI to emerge between two crossing field-dressed PESs.
This rotation angle together with the internuclear separation span the two-dimensional light-induced
BS along which degeneracy is lifted to first order around the LICI due to the dressing field.

Investigating light-induced nonadiabatic effects in polyatomic
molecules is expected to be more challenging than in diatomics.
On the one hand, the existence of several vibrational degrees of freedom enables a two-dimensional
BS to be defined without further involving any rotational coordinate.
On the other hand, natural CIs are so ubiquitous in polyatomics that their occurrence,
with subsequent natural nonadiabatic effects, are likely to be strongly mixed with the
light-induced ones.
It is thus desirable to consider situations where a clear separation and identification of the
effects of natural and light-induced conical intersections can be made,
so as to clarify and understand the effect solely caused by the LICI.

Experimental works\cite{Buksbaum,Banares}
have already invoked the concept of a LICI to provide qualitative interpretation to the results of
laser-induced isomerization and photodissociation in polyatomic molecules.
A general theory on LICIs in polyatomics discusses strategies of its exploitation and application in such systems.\cite{Lenz5}
The present study goes beyond previous investigations and makes an attempt to identify some ``direct observable signature'' of LICIs in polyatomic molecules.

It is well established in the field of ``natural'' nonadiabatic molecular phenomena that
the breakdown of the BO approximation induces strong mixing between
the vibrational levels within coupled adiabatic PESs.
The Franck--Condon approximation is no longer valid,
which typically leads at low energies to the so-called ``intensity borrowing'' effect.\cite{Lenz1}
This is a characteristic fingerprint of nonadiabatic effects in molecular spectroscopy,
manifested by irregular variations of spectral peak intensities and the appearance of unexpected levels.

Along this line, our current work presents strong intensity-borrowing effects, now appearing
in the field-dressed low-energy vibronic spectrum of the H$_{2}$CO (formaldehyde) molecule
and absent in the absence of the field,
providing direct evidence for the LICI and its effect.
The photochemistry of H$_{2}$CO has been studied in various contexts and the corresponding literature is abundant and beyond the scope of the present work.
We chose H$_{2}$CO for two reasons: first, because there is no natural CI in the vicinity of the Franck--Condon region.
A seam of CIs has been characterized for example by Araujo et al.\cite{ara08:7489, ara09:144301, ara10:12016}
but it is protected by a transition barrier at low energies;
second, this system presents the great advantage of not having any first-order nonadiabatic coupling
between the ground and first singlet excited electronic states around its equilibrium geometry.
As a consequence, such a situation allows unambiguous identification of light-induced nonadiabatic
phenomena, clearly separated from other ``natural'' nonadiabatic effects.

By using a two-step protocol we simulate the weak-field absorption
and stimulated emission spectra of the \emph{field-dressed} H$_{2}$CO molecule at low energy (infrared domain)
taking into account all six vibrational degrees of freedom explicitly.
First, we compute the field-dressed states which are superpositions of
field-free molecular eigenstates coupled by a medium-intensity laser dressing pump pulse switched on adiabatically.
Second, we assume that the low-energy spectrum of the field-dressed molecule can be accessed
with a weak-intensity steady-state probe. The dipole transition amplitudes between the
field-dressed states are thus evaluated within the typical framework of first-order time-dependent
perturbation theory, revealing intensity borrowing in the infrared domain for the field-dressed molecule.

\section{Theory and computational protocol}

\subsection{Working Hamiltonian}
Let us start with describing the working Hamiltonian and the protocol used
to compute field-dressed states and field-dressed spectra.
Throughout the current work the two singlet electronic states
$\textrm{S}_0 ~ (\tilde{\textrm{X}} ~ ^1\textrm{A}_1)$
and $\textrm{S}_1 ~ (\tilde{\textrm{A}} ~ ^1\textrm{A}_2)$ of H$_{2}$CO are
taken into account, the corresponding six-dimensional PESs are denoted by
$V_{\textrm{X}}$ and $V_{\textrm{A}}$, respectively.
We assume that the electronic states X and A are
coupled by a time-dependent external electric field 
$\mathbf{E}(t) = \mathbf{E}_0 \cos(\omega t)$.
In the static Floquet-state representation\cite{Floquet1,Floquet2} the
Hamiltonian matrix has the following form
\begin{equation}
    \hat{H}=
       \begin{bmatrix}\hat{T} & 0\\
                            0 & \hat{T}
       \end{bmatrix}+
       \begin{bmatrix}
                      V_{\textrm{X}}+n\hbar\omega & W\\
                      W & V_{\textrm{A}}+(n-1)\hbar\omega
       \end{bmatrix}
  \label{eq:H_rwa}
\end{equation}
where $\hat{T}$ is the vibrational kinetic energy operator constructed in a
fully exact and numerical way\cite{09MaCzCs,11FaMaCs} and $n$ is the Fourier index that labels the
different light-induced potential channels. The operator $W=-\mathbf{d}(\mathbf{R}) \mathbf{e} E_{0}/2$
describes the coupling between the electronic states X and A 
with $\mathbf{d}(\mathbf{R})$ being the body-fixed molecular transition dipole moment (TDM)
vector depending on the vibrational coordinates $\mathbf{R}$.
Moreover, $\mathbf{e}$, $E_{0}$ and $\omega$ denote the polarization, amplitude and angular frequency
of the dressing laser field, respectively. We used linearly-polarized electric fields throughout
this work. The rotational degrees of freedom were omitted from our computational protocol and
the orientation of the molecule was fixed with respect to the external electric field.
Finally, we note that in all practical computations we went beyond the approximate two-by-two
Floquet Hamiltonian of eqn \eqref{eq:H_rwa} and used the following light-dressed Hamiltonian
\begin{equation}
    \hat{H}=
       \begin{bmatrix}
       			   \hat{H}_{-n_\textrm{max}} & \hat{D} & 0 & 0 & 0 & 0 & 0 \\
			   \hat{D}^\dag & \ddots & \vdots & \vdots & \vdots & \reflectbox{$\ddots$} & 0 \\
			   0 & \cdots & \hat{H}_{-1} & \hat{D} & 0 & \cdots & 0 \\
			   0 & \cdots & \hat{D}^\dag & \hat{H}_{0} & \hat{D} & \cdots & 0 \\
			   0 & \cdots & 0 & \hat{D}^\dag & \hat{H}_{1} & \cdots & 0 \\
			   0 &  \reflectbox{$\ddots$} &\vdots & \vdots & \vdots & \ddots & \hat{D} \\
			   0 & 0 & 0 & 0 & 0 & \hat{D}^\dag & \hat{H}_{n_\textrm{max}}
       \end{bmatrix}
  \label{eq:H_exact}
\end{equation}
with
\begin{equation}
    \hat{H}_n=
       \begin{bmatrix}\hat{T}+V_{\textrm{X}}+n\hbar\omega & 0\\
                            0 & \hat{T}+V_{\textrm{A}}+n\hbar\omega
       \end{bmatrix}
\end{equation}
and
\begin{equation}
    \hat{D}=
       \begin{bmatrix} W_{\textrm{X}} & W\\
                             W & W_{\textrm{A}}
       \end{bmatrix}
\end{equation}
where the terms $W_{\textrm{X}}=-\mathbf{d}_{\textrm{X}}(\mathbf{R}) \mathbf{e} E_{0}/2$ and
$W_{\textrm{A}}=-\mathbf{d}_{\textrm{A}}(\mathbf{R}) \mathbf{e} E_{0}/2$ describe the
interaction between the external electric field and the permanent dipole moments (PDMs)
of the electronic states X and A.\cite{19ToCsHa}
In what follows, for the sake of simplicity, the discussion will be limited to the
two-by-two Floquet Hamiltonian of eqn \eqref{eq:H_rwa}.

After diagonalizing the potential energy matrix of eqn \eqref{eq:H_rwa}
one can obtain the adiabatic potential energy surfaces $V_{\textrm{lower}}$
and $V_{\textrm{upper}}$. These two PESs can cross
each other, giving rise to a LICI, whenever the conditions $W=0$
and $V_{\textrm{X}}=V_{\textrm{A}}-\hbar\omega$ are simultaneously fulfilled.

The field-dressed eigenfunctions $|\Phi_{k}(n)\rangle$
and quasienergies $\epsilon_{k}(n)$ can be obtained by determining the eigenpairs
of the Hamiltonian of eqn \eqref{eq:H_rwa}. The eigenfunctions $|\Phi_{k}(n)\rangle$
can be expanded as the linear combination of products of field-free
molecular vibronic eigenstates (denoted by $|\textrm{X}i\rangle$ and $|\textrm{A}i\rangle$
for the electronic states $\textrm{X}$ and $\textrm{A}$, respectively)
and the Fourier vectors of the Floquet states, that is 
\begin{equation}
    |\Phi_{k}(n)\rangle=\sum_{i}C_{\textrm{X}i}^{(k)}|\textrm{X}i\rangle|n\rangle+\sum_{i}C_{\textrm{A}i}^{(k)}|\textrm{A}i\rangle|n-1\rangle.
  \label{eq:dressed}
\end{equation}
In eqn \eqref{eq:dressed} $i$ labels vibrational eigenstates and $|n\rangle$
is the $n$th Fourier vector of the Floquet state. The expansion coefficients
$C_{\textrm{X}i}^{(k)}$ and $C_{\textrm{A}i}^{(k)}$
can be obtained by diagonalizing the light-dressed Hamiltonian of eqn \eqref{eq:H_rwa}
after constructing its matrix representation.

To move forward, we briefly discuss the spectroscopy of field-dressed molecules.\cite{Gabi11}
Namely, we compute transition amplitudes between field-dressed
states $|\Phi_{k}(n)\rangle$ and $|\Phi_{l}(m)\rangle$ assuming a
weak probe pulse which allows us to use first-order time-dependent perturbation
theory for the evaluation of transition amplitudes following
the standard procedure of theoretical molecular spectroscopy.\cite{Bunker}
The transition amplitudes are the matrix elements of the
electric dipole moment operator $\hat{d}_{\alpha}$ ($\alpha=x,y,z$)
between two field-dressed states of eqn \eqref{eq:dressed},
\begin{equation}
    \begin{split}
        & \langle\Phi_{k}(n)|\hat{d}_{\alpha}|\Phi_{l}(n)\rangle=\\
        & \sum_{i}\sum_{j}C_{\textrm{X}i}^{(k)*}C_{\textrm{X}j}^{(l)}\langle\textrm{X}i|\hat{d}_{\alpha}|\textrm{X}j\rangle+
          \sum_{i}\sum_{j}C_{\textrm{A}i}^{(k)*}C_{\textrm{A}j}^{(l)}\langle\textrm{A}i|\hat{d}_{\alpha}|\textrm{A}j\rangle
    \end{split}
  \label{eq:intprd}
\end{equation}
and
\begin{equation}
    \begin{split}
        &\langle\Phi_{k}(n)|\hat{d}_{\alpha}|\Phi_{l}(m)\rangle= \\
        &\delta_{m,n+1}\sum_{i}\sum_{j}C_{\textrm{X}i}^{(k)*}C_{\textrm{A}j}^{(l)}\langle\textrm{X}i|\hat{d}_{\alpha}|\textrm{A}j\rangle+ \\
        &\delta_{m,n-1}\sum_{i}\sum_{j}C_{\textrm{A}i}^{(k)*}C_{\textrm{X}j}^{(l)}\langle\textrm{A}i|\hat{d}_{\alpha}|\textrm{X}j\rangle
    \end{split}
  \label{eq:inttrd}
\end{equation}
where $n\ne m$. One can notice that while transitions $|\Phi_{k}(n)\rangle\rightarrow|\Phi_{l}(n)\rangle$
are associated with the PDMs related to the electronic
states X and A, the transitions $|\Phi_{k}(n)\rangle\rightarrow|\Phi_{l}(n+1)\rangle$
and $|\Phi_{k}(n)\rangle\rightarrow|\Phi_{l}(n-1)\rangle$ are governed
by the TDM. The corresponding transition energies
are $\epsilon_{l}(n)-\epsilon_{k}(n)$, $\epsilon_{l}(n+1)-\epsilon_{k}(n)=\epsilon_{l}(n)-\epsilon_{k}(n)+\hbar\omega$
and $\epsilon_{l}(n-1)-\epsilon_{k}(n)=\epsilon_{l}(n)-\epsilon_{k}(n)-\hbar\omega$,
respectively. Eqn \eqref{eq:intprd} serves as our working formula for
the computation of the low-energy part of the vibronic spectrum.
The expressions of eqn \eqref{eq:intprd} and eqn \eqref{eq:inttrd} pertain to both the field-dressed absorption and
stimulated emission processes.
Peaks associated with absorption and stimulated emission are separated by comparing
the quasienergies of the initial and final field-dressed states.
The intensity of transitions between field-dressed states can be obtained as 
\begin{equation}
    I_{kl}\propto\omega_{kl}\sum_{\alpha=x,y,z}|\langle\Phi_{k}(n)|\hat{d}_{\alpha}|\Phi_{l}(m)\rangle|^{2}
  \label{eq:intensity}
\end{equation}
where $\omega_{kl}$ denotes the transition frequency.

\subsection{The H$_{2}$CO molecule and technical details}
\label{sec:technical}
\begin{figure}[hbt!]
 \centering
   \includegraphics[scale=0.3]{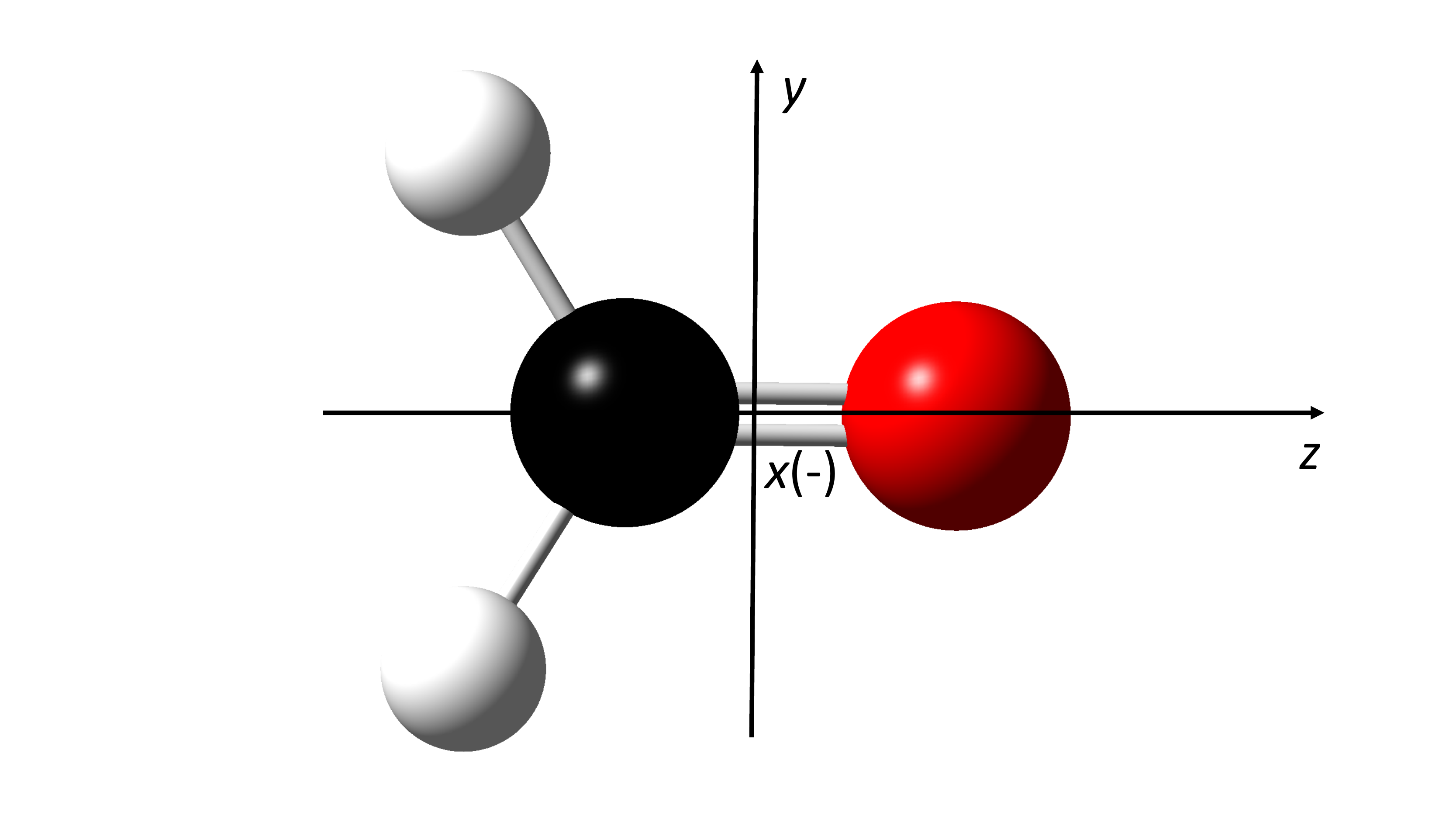}
   \caption{Equilibrium structure of the H$_{2}$CO molecule in its electronic ground state (X)
   		and the body-fixed coordinate system chosen (the $x$ axis is directed inwards, as indicated by the - sign).}
   \label{fig:structure}
\end{figure}

\begin{figure}[hbt!]
 \centering
   \includegraphics[scale=0.75]{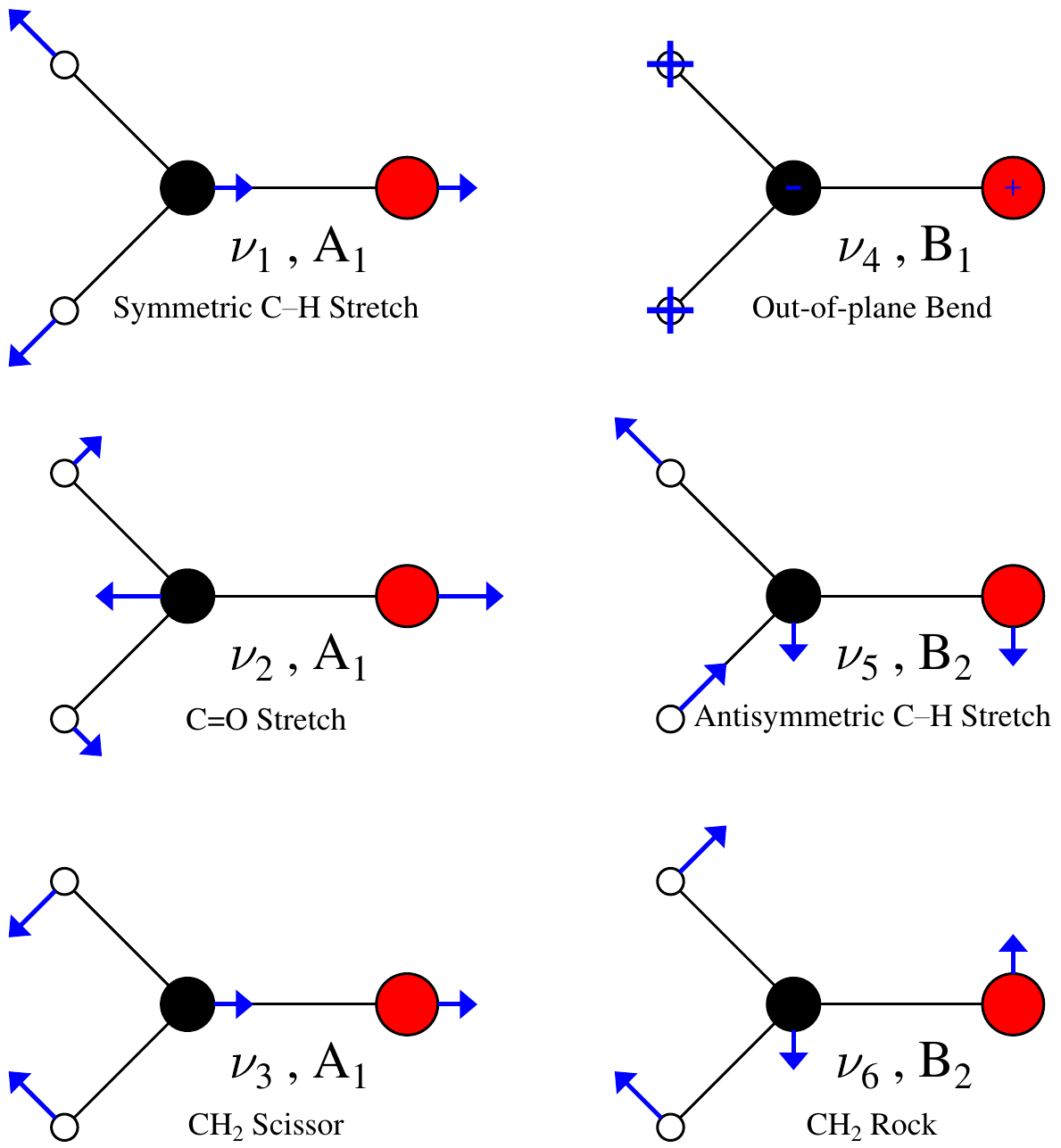}
   \caption{Normal modes of the H$_{2}$CO molecule.}
   \label{fig:normalmodes}
\end{figure}

The $V_{\textrm{X}}$ ($\textrm{S}_0$ electronic state, X) and $V_{\textrm{A}}$ ($\textrm{S}_1$ electronic state, A) PESs
were taken from refs. \citenum{Bowman2} and \citenum{Bowman1}, respectively,
the corresponding vertical and adiabatic excitation energies are $28954 ~ \textrm{cm}^{-1}$ and $28138 ~ \textrm{cm}^{-1}$.
The planar ground-state (X) equilibrium structure of H$_{2}$CO
($C_{2v}$ point-group symmetry) is shown in Fig. \ref{fig:structure} while
its normal modes are summarized in Fig. \ref{fig:normalmodes} and Table \ref{tbl:normalmodes}.
Note that the definition of the Cartesian axes in Fig. \ref{fig:structure} is in line with the Mulliken convention.\cite{55Mulliken}
The excited electronic state (A) has a double-well structure along the out-of-plane ($\nu_4$) mode
and the two equivalent nonplanar equilibrium structures are connected by a planar transition state structure,
the respective barrier height is $315 ~ \textrm{cm}^{-1}$.

\begin{table}[hbt!]
   \caption{\ Normal mode labels, $C_{2v}$ irreducible representations, description of normal modes and anharmonic fundamentals
              (obtained by 6D variational computations in the electronic ground state (X)
               of H$_{2}$CO, in units of $\textrm{cm}^{-1}$).}
   \label{tbl:normalmodes}
   \begin{center}
   \begin{tabular}{cccc}
        mode & symmetry & description & $\omega / \textrm{cm}^{-1}$ \\
        \hline
        $\nu_1$ & $\textrm{A}_1$ & sym C-H stretch & $2728.4$ \\
        $\nu_2$ & $\textrm{A}_1$ & C=O stretch & $1738.1$ \\
        $\nu_3$ & $\textrm{A}_1$ & CH$_2$ scissor & $1466.0$ \\
        $\nu_4$ & $\textrm{B}_1$ & out-of-plane bend & $1147.0$ \\
        $\nu_5$ & $\textrm{B}_2$ & antisym C-H stretch & $2819.9$ \\
        $\nu_6$ & $\textrm{B}_2$ & CH$_2$ rock & $1234.5$
   \end{tabular}
   \end{center}
\end{table}

The six-dimensional vibrational Schr\"odinger equation was solved variationally by the numerically exact and general
rovibrational code GENIUSH\cite{09MaCzCs,11FaMaCs,12CsFaSz} for both PESs.
The body-fixed Cartesian position vectors of the nuclei were parameterized using polyspherical coordinates\cite{92ChIu} and
the body-fixed axes were oriented according to Eckart conditions\cite{35Eckart} using
the equilibrium structure of the X electronic state as reference structure.
A symmetry-adapted six-dimensional direct-product discrete variable representation (DVR) basis and atomic mass values
$m_\textrm{C} = 12.0 \, \textrm{u}$, $m_\textrm{O} = 15.994915 \, \textrm{u}$ and $m_\textrm{H} = 1.007825 \, \textrm{u}$
were employed throughout the nuclear motion computations.
The vibrational Hamiltonian matrix was separated into four blocks corresponding to the four irreducible representations
of the $S_2^*$ molecular symmetry group\cite{Bunker} which is isomorphic to the $C_{2v}$ point group.
In order to assist the interpretation of the field-dressed spectra presented in Section \ref{sec:results}, the vibrational
eigenstates of the A electronic state were recomputed using the DEWE program package\cite{07MaCzSu,11FaMaFu}
and the rectilinear normal coordinates corresponding to the planar transition state structure of the A electronic state.
Then, one-dimensional wave function cuts along the normal coordinates were evaluated and the nodal structure of
these one-dimensional cuts was inspected to assign the eigenstates with vibrational quantum numbers.

The permanent and transition dipole moment, i.e., PDM and TDM, surfaces
required by the computation of the field-dressed states and transition amplitudes,
were generated by Taylor expansions up to second order
using the polyspherical coordinates employed by the vibrational eigenstate computations.
The Taylor series are centered at the equilibrium structure of the X electronic state (TDM and X-state PDM)
and at the transition state structure of the A electronic state (A-state PDM), respectively.
The PDM and TDM components were referenced in the Eckart frame described above and the necessary
dipole derivatives were evaluated numerically at the CAM-B3LYP/6-31G* level of theory.
The symmetry properties of the body-fixed PDM and TDM components are summarized
in Table \ref{tbl:dipsymm}.
The field-dressed states were computed by diagonalizing the light-dressed Hamiltonian
of eqn \eqref{eq:H_exact} in the direct-product basis of field-free molecular eigenstates and Fourier vectors
of the Floquet states $|n\rangle$, the maximal Fourier component was set to $n_\textrm{max}=2$ which
was sufficient for obtaining converged results.

\begin{table}[hbt!]
   \caption{\ Symmetry properties ($C_{2v}$ irreducible representations) of the body-fixed components of
              the permanent (PDM) and transition (TDM) dipole moments.}
   \label{tbl:dipsymm}
   \begin{center}
   \begin{tabular}{ccc}
         & PDM & TDM \\
        \hline
        $x$ & $\textrm{B}_1$ & $\textrm{B}_2$ \\
        $y$ & $\textrm{B}_2$ & $\textrm{B}_1$ \\
        $z$ & $\textrm{A}_1$ & $\textrm{A}_2$ 
   \end{tabular}
   \end{center}
\end{table}

The symmetries of the X and A electronic states are $\textrm{A}_1$ and $\textrm{A}_2$ at the
Franck--Condon point of $C_{2v}$ symmetry.
According to Table \ref{tbl:normalmodes}, H$_{2}$CO does not have any vibrational modes of $\textrm{A}_2$
symmetry, and therefore, there is no vibration available for the expansion of the first-order
nonadiabatic coupling of $\textrm{A}_2$ symmetry at the Franck--Condon point.
In addition, all components of the TDM vanish at the Franck--Condon point due to symmetry and hence
all natural and light-induced couplings are zero at $C_{2v}$ geometries.
The $x$-component of the TDM is a linear function of both $\textrm{B}_2$ modes, while
the $y$-component is a linear function of the $\textrm{B}_1$ mode,
which can induce LICIs when distorting the geometry for a given field polarization.

\subsection{Dressing mechanism and light-induced conical intersections}
\label{sec:lici}
In what follows, a simple model facilitating the analysis of the field-dressed
spectra presented in Section \ref{sec:results} is outlined.
We assume that the dressing field is turned on adiabatically and the initial field-dressed state
$| \Phi_i(n) \rangle$ is chosen as the field-dressed state which gives maximal overlap with the
vibrational ground state of the X electronic state.
If the dressing field couples the vibrational ground state of X
to $N$ vibrational eigenstates of A, the initial field-dressed state becomes
\begin{equation}
    | \Phi_i(n) \rangle = 
        C_{\textrm{X}0}^{(i)} |\textrm{X}0\rangle |n\rangle +
        \sum_{k=1}^{N} C_{\textrm{A}k}^{(i)} |\textrm{A}k\rangle |n-1\rangle.
    \label{eq:initial_state}
\end{equation}

\begin{figure}[hbt!]
 \centering
   \includegraphics[scale=0.75]{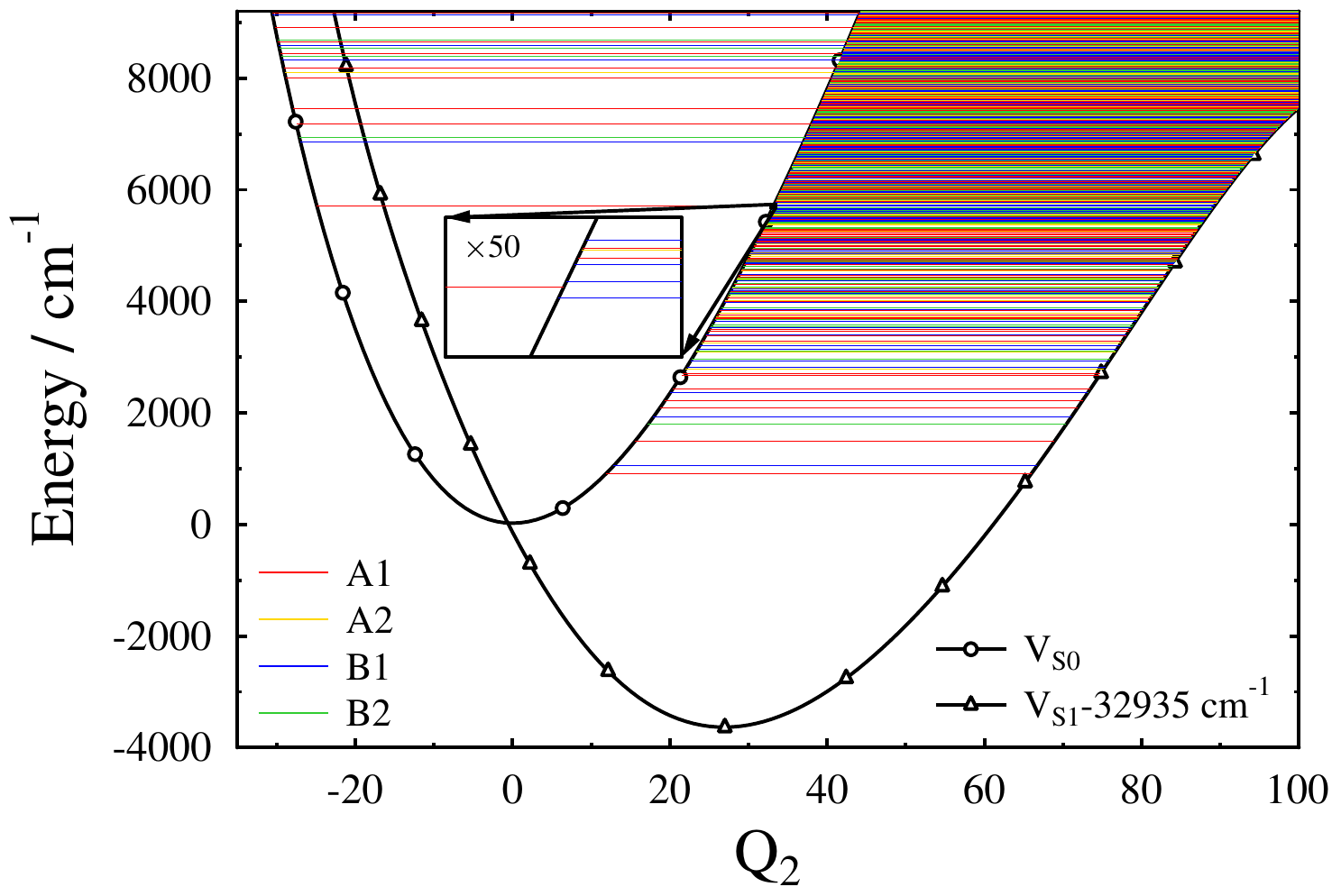}
   \caption{One-dimensional field-free potential energy cuts along the $\nu_2$ (C=O stretch) normal mode.
            The excited-state potential curve ($V_\textrm{S1}$ in the figure) is shifted down
            by the photon energy value corresponding to $32935 ~ \textrm{cm}^{-1}$.
            Vibrational energy levels are indicated by horizontal lines with colours encoding
            irreducible representations of the $C_{2v}$ point group.}
   \label{fig:pes_1D}
\end{figure}

\begin{figure}[hbt!]
 \centering
   \includegraphics[scale=0.8]{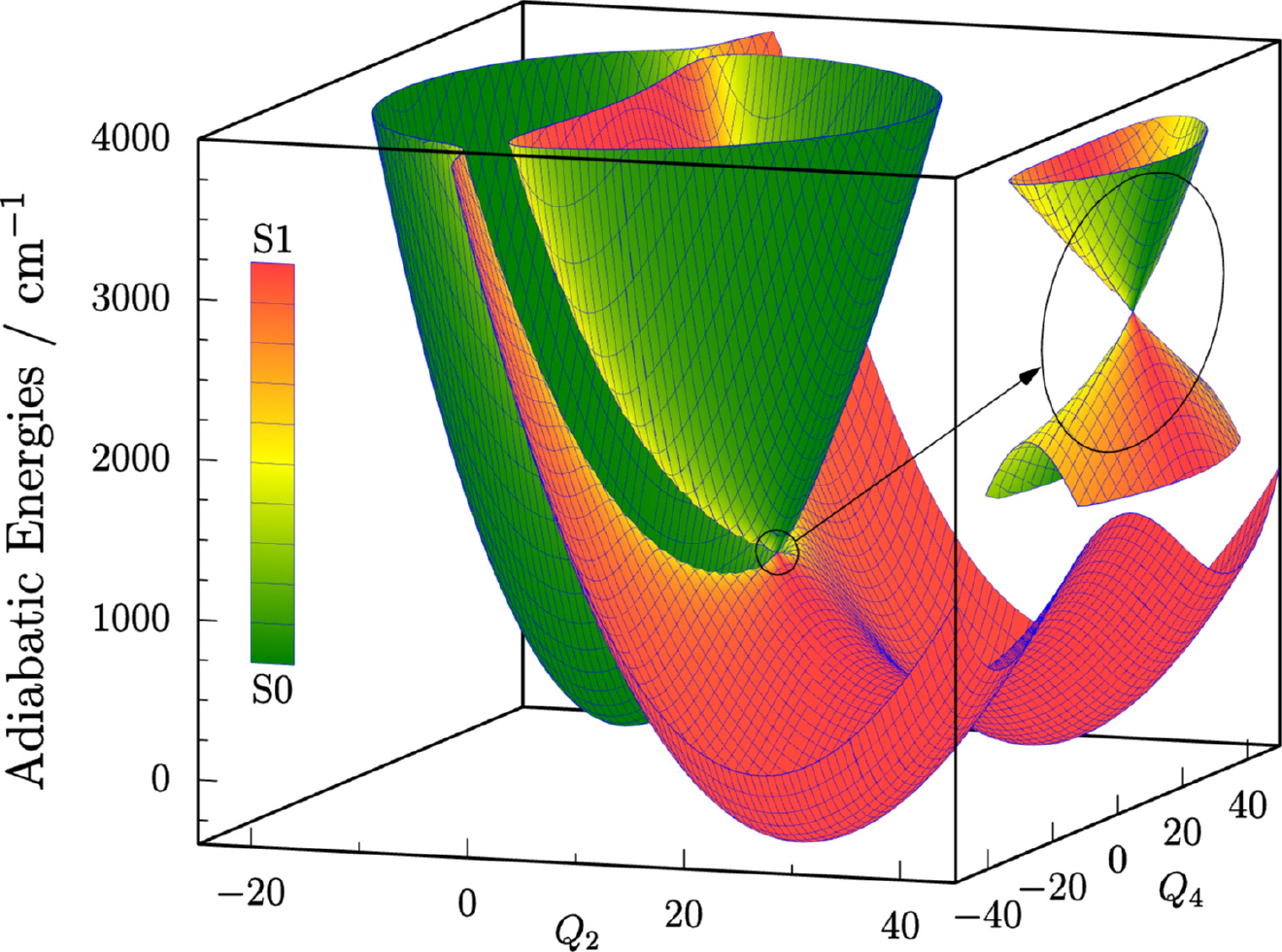}
   \caption{Two-dimensional light-induced adiabatic potential energy surfaces along
            the $\nu_2$ (C=O stretch) and $\nu_4$ (out-of-plane bend) normal modes.
            The dressing wavenumber and intensity are chosen as $\omega_\textrm{d} = 29000 ~ \textrm{cm}^{-1}$
            and $I_\textrm{d} = 10^{14} ~ \textrm{W}/\textrm{cm}^{2}$, respectively, and the dressing
            field is polarized along the body-fixed $y$ axis.
            This intensity is chosen for a better visualization only.
            The light-induced conical intersection is highlighted in the inset on the right-hand side of the
            figure. The character of the adiabatic potential energy surfaces is indicated by
            different colours (see the legend on the left).
            }
   \label{fig:pes_2D}
\end{figure}

Such a situation is featured in Fig. \ref{fig:pes_1D} where one-dimensional cuts of the
$V_{\textrm{X}}$ and $V_{\textrm{A}}$ PESs are shown along the $\nu_2$ normal mode
(normal coordinates other than $Q_2$ are set to zero).
In Fig. \ref{fig:pes_1D}, the system is dressed with photons corresponding to 
$\omega = 32935 ~ \textrm{cm}^{-1}$ and the $V_{\textrm{A}}$ PES is
shifted down with the energy of the dressing photon. Consequently, the vibrational
ground state of X becomes nearly resonant with multiple close-lying excited vibrational
eigenstates of A and field-dressed states of eqn \eqref{eq:initial_state} are formed.
Fig. \ref{fig:pes_2D} provides two-dimensional cuts of the resulting light-induced
adiabatic PESs with the dressing wavenumber of $\omega_\textrm{d} = 29000 ~ \textrm{cm}^{-1}$
and dressing intensity of $I_\textrm{d} = 10^{14} ~ \textrm{W}/\textrm{cm}^{2}$
(the dressing field is polarized along the body-fixed $y$ axis).
The two-dimensional cuts, given as functions of the $Q_2$ and $Q_4$ normal coordinates
(the remaining four normal coordinates are set to zero),
clearly show that a LICI is formed between the two light-induced adiabatic PESs.

If the final field-dressed state remains an eigenstate of the field-free molecule, that is,
\begin{equation}
    | \Phi_j(n) \rangle = |\textrm{X}j\rangle |n\rangle
    \label{eq:final_state1}
\end{equation}
or
\begin{equation}
    | \Phi_{j'}(n) \rangle = |\textrm{A}j'\rangle |n-1\rangle,
    \label{eq:final_state2}
\end{equation}
then the transition amplitudes for the transitions $| \Phi_i(n) \rangle \rightarrow | \Phi_j(n) \rangle$
and $| \Phi_i(n) \rangle \rightarrow | \Phi_{j'}(n) \rangle$ take the form
\begin{equation}
    \langle \Phi_{j}(n) | \hat{d}_\alpha | \Phi_i(n) \rangle = 
        C_{\textrm{X}0}^{(i)} \langle \textrm{X}j| \hat{d}_\alpha |\textrm{X}0 \rangle
    \label{eq:transition_amplitude1}
\end{equation}
and
\begin{equation}
    \langle \Phi_{j'}(n) | \hat{d}_\alpha | \Phi_i(n) \rangle = 
        \sum_{k=1}^{N} C_{\textrm{A}k}^{(i)} \langle \textrm{A}j' | \hat{d}_\alpha | \textrm{A}k\rangle.
    \label{eq:transition_amplitude2}
\end{equation}
The physical interpretation of the previous equations is outlined as follows.
Peaks that correspond to the transitions $|\textrm{X}0\rangle \rightarrow |\textrm{X}j\rangle$ of the
field-free vibrational spectrum appear also in the field-dressed spectrum, but their intensities
in the field-dressed case are proportional to
$\lvert C_{\textrm{X}0}^{(i)} \langle \textrm{X}j| \hat{d}_\alpha |\textrm{X}0 \rangle \rvert^2$
(see eqn \eqref{eq:transition_amplitude1}), falling short of the corresponding field-free intensity values
since $\lvert C_{\textrm{X}0}^{(i)} \rvert^2 < 1$.
The second group of peaks (see eqn \eqref{eq:transition_amplitude2}) appear only in the field-dressed case as a consequence of
the field-induced couplings between the vibrational ground state of X and vibrational eigenstates of A.
As we will see in Section \ref{sec:results}, these peaks appear primarily in the lower half of the
studied region of the field-dressed spectrum and they are associated with the PDM
of the A electronic state according to eqn \eqref{eq:transition_amplitude2}.

\section{Results and discussion}
\label{sec:results}
\begin{figure}[hbt!]
 \centering
   \includegraphics[scale=0.65]{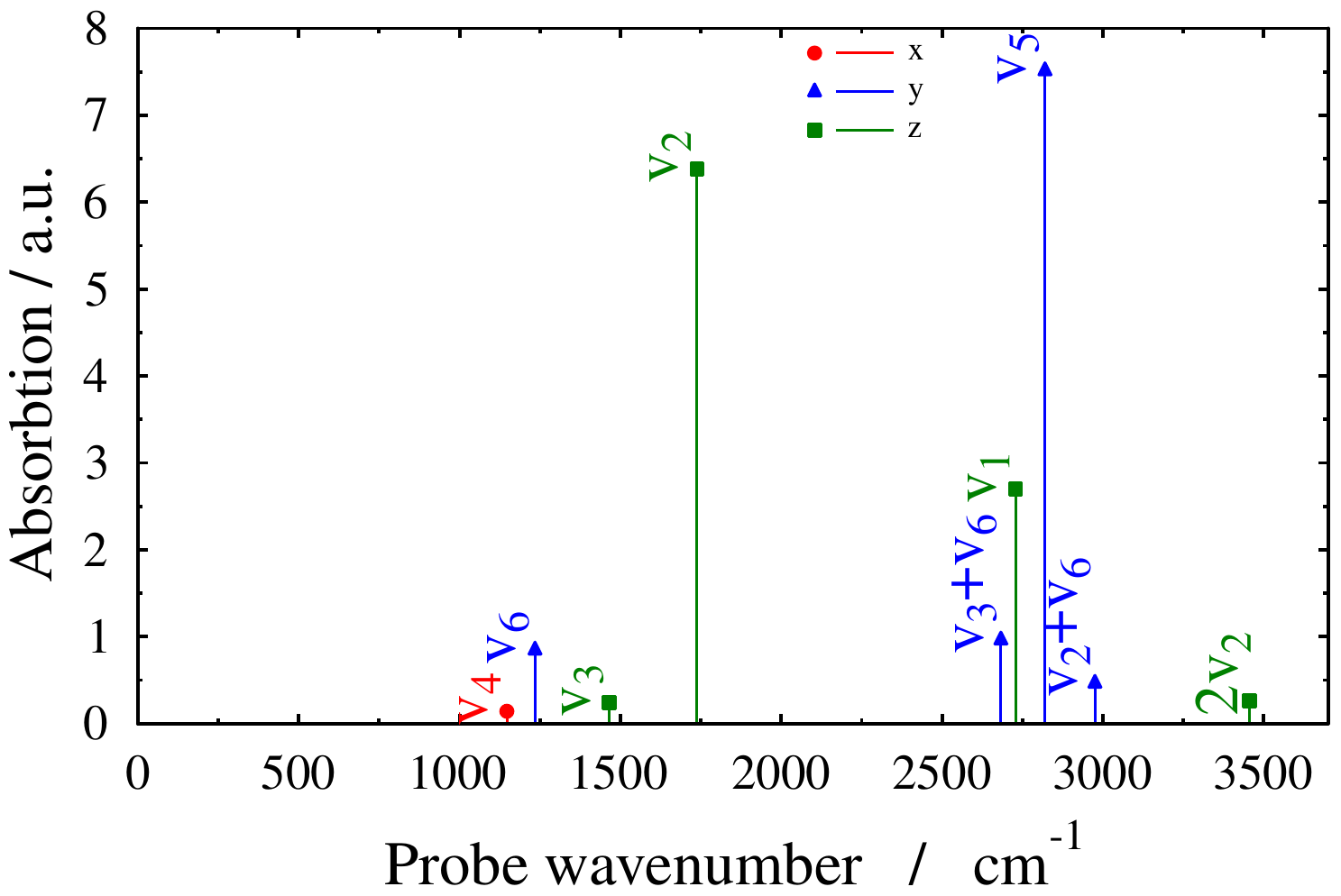}
   \caption{Field-free vibrational spectrum of the H$_2$CO molecule in its electronic ground state (X).
            The different peaks correspond to transitions from the vibrational ground state to
            vibrationally-excited eigenstates as indicated by the normal mode labels.
            Transitions polarized along the $x$, $y$ and $z$ axes are shown by the markers
            $\bullet$, $\blacktriangle$ and $\blacksquare$, respectively.}
   \label{fig:spectrum_ff}
\end{figure}

Let us start with the field-free vibrational spectrum of H$_2$CO in its electronic ground
state X, shown in Fig. \ref{fig:spectrum_ff}.
The spectrum in Fig. \ref{fig:spectrum_ff} was
computed using the six-dimensional variational vibrational eigenstates provided by
GENIUSH\cite{09MaCzCs,11FaMaCs,12CsFaSz} and the PDM surface of the X electronic state described in
Section \ref{sec:technical}.
The field-free vibrational spectrum has been found to agree well with the anharmonic
vibrational spectrum computed by the Gaussian 09 program package.\cite{g09}
As expected, the field-free vibrational spectrum consists of a moderate number of peaks, corresponding
to vibrational transitions from the initially populated vibrational ground state to vibrationally-excited eigenstates.
Besides the six fundamental transitions (denoted with the normal mode labels
$\nu_i$ in Fig. \ref{fig:spectrum_ff}) two combination transitions
($\nu_2+\nu_6$ and $\nu_3+\nu_6$) and an overtone transition ($2\nu_2$)
carry appreciable intensity. It is clearly visible in Fig. \ref{fig:spectrum_ff}
that no peaks appear below $1100 ~ \textrm{cm}^{-1}$ in the field-free vibrational spectrum.

\begin{figure}[hbt!]
 \centering
   \includegraphics[scale=0.65]{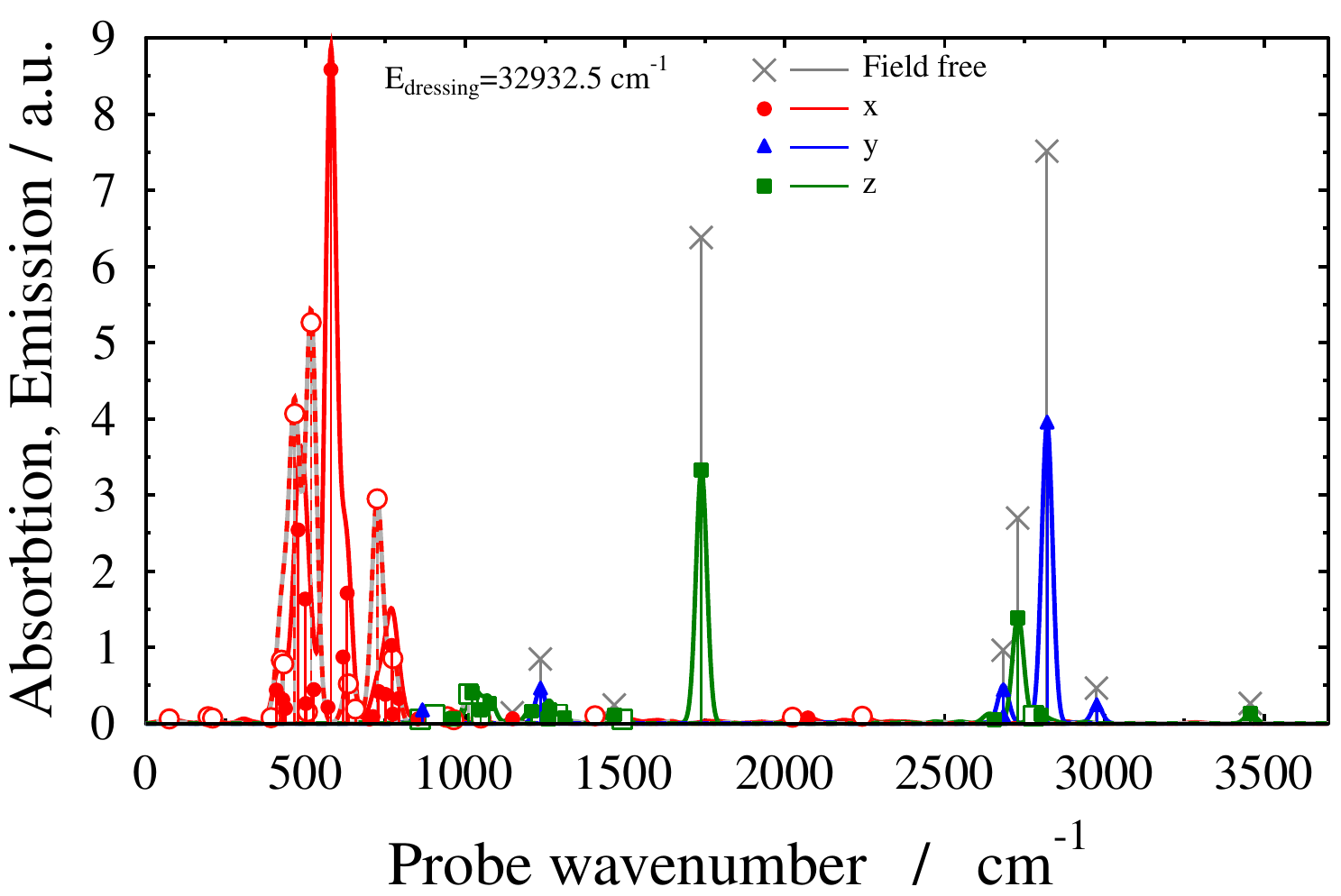}
   \includegraphics[scale=0.65]{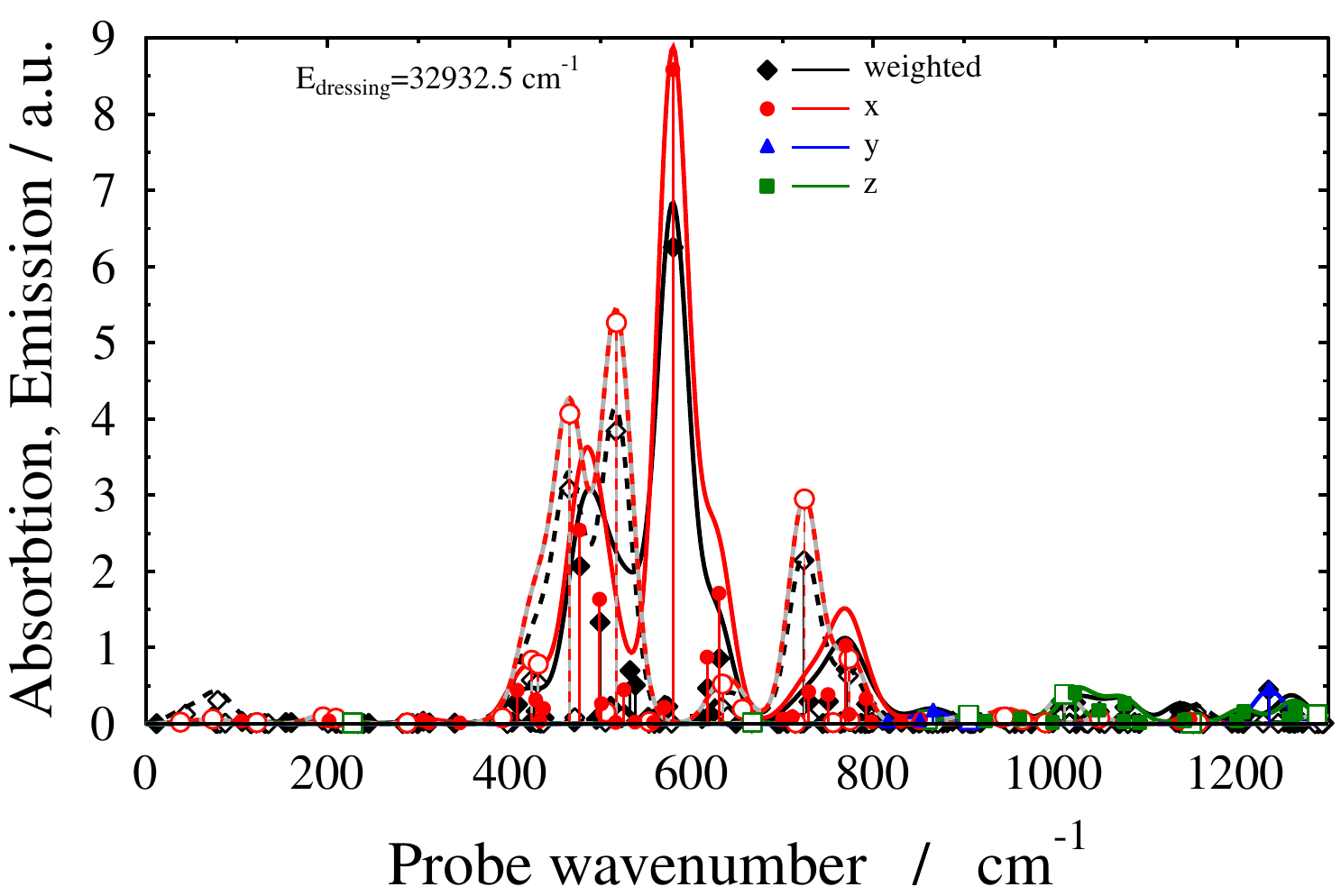}
   \caption{Absorption (solid line, full markers) and stimulated emission (dashed line, empty markers) spectra of H$_{2}$CO
            dressed with a laser field linearly polarized along the
            body-fixed $y$ axis with $I_\textrm{d} = 10^{11} ~ \textrm{W}/\textrm{cm}^2$
            and $\omega_\textrm{d} = 32932.5 ~ \textrm{cm}^{-1}$.
            Transitions polarized along the $x$, $y$ and $z$ axes are shown
            by the markers $\bullet$, $\blacktriangle$ and $\blacksquare$, respectively.
            The field-dressed spectrum is compared to the field-free vibrational spectrum, denoted by
            the marker $\times$ (upper panel).
            The weighted field-dressed spectrum (see text) is indicated by the marker \ding{117}. 
            The lower panel highlights the spectral range $0-1200 ~ \textrm{cm}^{-1}$,
            most relevant for the analysis of field-dressed spectra.
            Stick spectra were convolved with Gaussian functions having a full
            width at half maximum of $40 ~ \textrm{cm}^{-1}$.
            The most salient feature of the field-dressed spectrum, as opposed to the field-free vibrational spectrum,
            is the appearance of numerous peaks below $1100 ~ \textrm{cm}^{-1}$.}
   \label{fig:spectrum_dressed1}
\end{figure}

We now turn to the discussion of the field-dressed spectra.
In the computations several values of the dressing intensity, ranging from
$I_\textrm{d} = 10^8 ~ \textrm{W}/\textrm{cm}^{2}$ to $I_\textrm{d} = 10^{11} ~ \textrm{W}/\textrm{cm}^2$,
and of the dressing wavenumber, ranging from $\omega_\textrm{d} = 28000 ~ \textrm{cm}^{-1}$ to
$\omega_\textrm{d} = 35000 ~ \textrm{cm}^{-1}$ (or from $\lambda_\textrm{d} = 357.14 ~ \textrm{nm}$
to $\lambda_\textrm{d} = 285.71 ~ \textrm{nm}$ in units of wavelength), were applied.
In the following we consider different dressing situations that lead to
striking effects in the field-dressed spectra.
Fig. \ref{fig:spectrum_dressed1} shows the absorption (solid line) and stimulated emission (dashed line)
spectra of H$_{2}$CO molecule dressed with a laser field
linearly polarized along the body-fixed $y$ axis with $I_\textrm{d} = 10^{11} ~ \textrm{W}/\textrm{cm}^2$
and $\omega_\textrm{d} = 32932.5 ~ \textrm{cm}^{-1}$ ($\lambda_\textrm{d} = 303.65 ~ \textrm{nm}$).
Two prominent features of the field-dressed spectrum in Fig. \ref{fig:spectrum_dressed1}
are the emergence of several peaks below $1100 ~ \textrm{cm}^{-1}$,
as opposed to the field-free vibrational spectrum, and the appearance of
induced emission peaks which are completely missing from the field-free vibrational spectrum.

The newly-emerging peaks, readily visible in the lower panel of Fig. \ref{fig:spectrum_dressed1},
can be classified into four main groups. The first two groups consist of peaks between
$400-650 ~ \textrm{cm}^{-1}$ ($\textrm{g}_1$) and $700-800 ~ \textrm{cm}^{-1}$ ($\textrm{g}_2$), while
the third and fourth groups involve peaks located in the intervals $800-900 ~ \textrm{cm}^{-1}$ ($\textrm{g}_3$)
and $1000-1100 ~ \textrm{cm}^{-1}$ ($\textrm{g}_4$).
The transitions of $\textrm{g}_1$ and $\textrm{g}_2$ types are all polarized along the body-fixed $x$ axis,
while the $\textrm{g}_3$ and $\textrm{g}_4$ transitions are polarized along the body-fixed $y$
(hardly discernible in Fig. \ref{fig:spectrum_dressed1})
and $z$ axes, respectively. As already discussed in Section \ref{sec:lici}, the peaks appearing
below $1100 ~ \textrm{cm}^{-1}$ in the field-dressed spectrum can be attributed
to admixtures of the vibrational eigenstates of the A electronic state
in the initial field-dressed state of eqn \eqref{eq:initial_state}. 
The initial field-dressed state in this particular case is a superposition of the
X vibrational ground state ($0.0 ~ \textrm{cm}^{-1}$, $\textrm{A}_1$ symmetry, $52.3\%$) and
two close-lying A vibrational eigenstates of $\textrm{B}_1$ symmetry, namely
$32931.2 ~ \textrm{cm}^{-1}$ ($36.0\%$) and $32937.2 ~ \textrm{cm}^{-1}$ ($11.5\%$)
where the numbers in the parentheses indicate the populations of the vibrational eigenstates in the initial field-dressed state.
These data were then used to compute weighted field-dressed spectra, defined 
as the weighted average of field-free spectra whose initial states correspond to
the X and A vibrational eigenstates that constitute the initial field-dressed state.
Since the weights are chosen as the populations of the vibrational eigenstates in the
initial field-dressed state, the weighted spectra are expected to be similar to the field-dressed
spectra. Nevertheless, the absolute square of the transition amplitude expression in eqn \eqref{eq:transition_amplitude2}
involves interference terms, therefore, the intensity patterns in the field-dressed spectra can differ from
the ones in the weighted spectra. 
We note that as the dressing field is polarized along the
body-fixed $y$ axis and the body-fixed $y$ component of the TDM transforms according to the $\textrm{B}_1$
irreducible representation, the X vibrational ground state ($\textrm{A}_1$ symmetry)
can be coupled only to vibrational eigenstates of $\textrm{B}_1$ symmetry of the A electronic state by the dressing field.
For all peaks shown in Fig. \ref{fig:spectrum_dressed1} the final field-dressed states
remain eigenstates of the field-free molecule to a very good approximation
and they can be written according to eqn \eqref{eq:final_state1} or eqn \eqref{eq:final_state2}.
Therefore, the considerations made below eqn \eqref{eq:transition_amplitude2} can be applied
to decipher the field-dressed spectrum shown in Fig. \ref{fig:spectrum_dressed1}.

\begin{figure}[hbt!]
 \centering
   \includegraphics[scale=0.65]{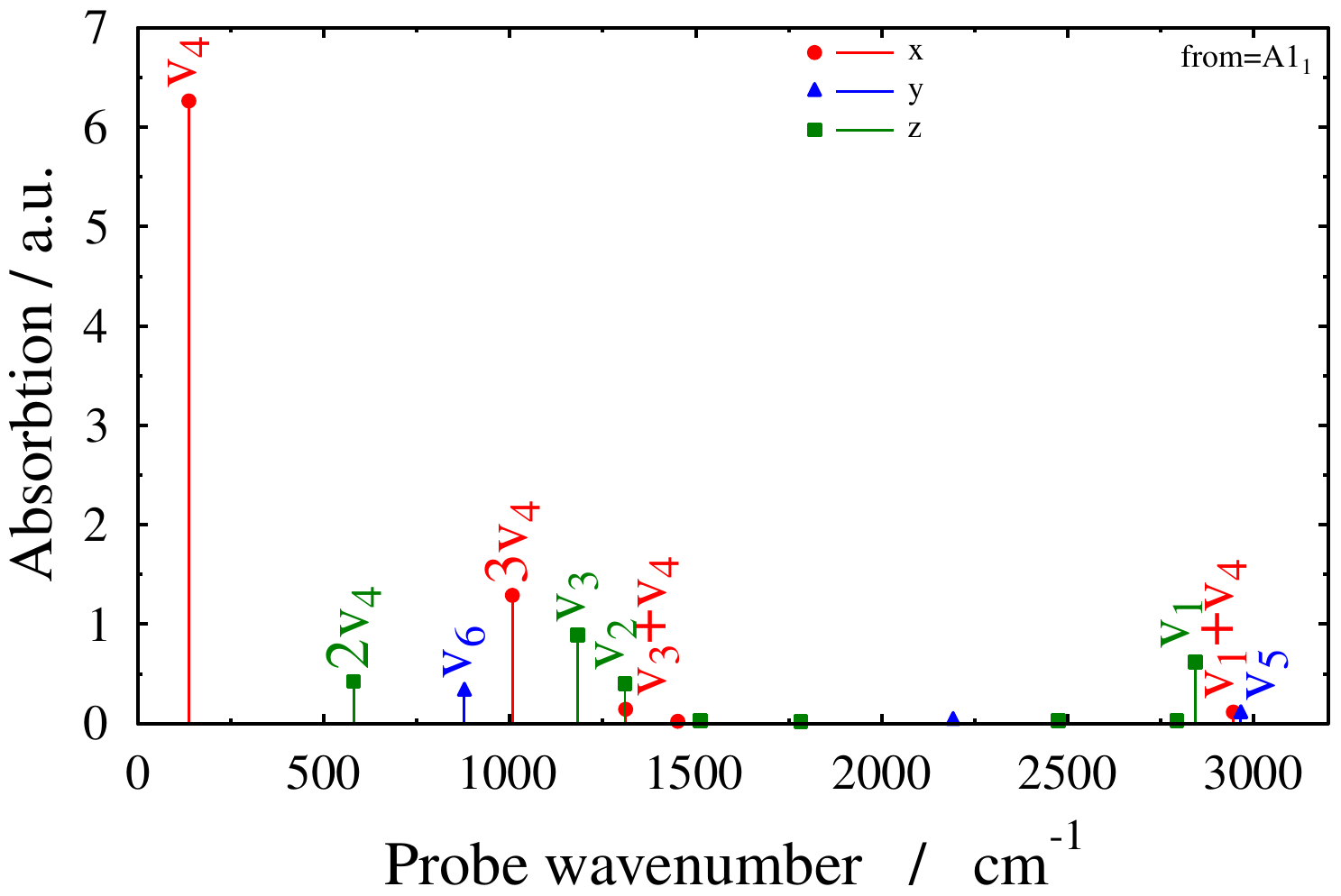}
   \caption{Field-free vibrational spectrum of the H$_2$CO molecule in its excited electronic state (A).
            The different peaks correspond to transitions from the vibrational ground state (GS) to
            vibrationally-excited eigenstates as indicated by the normal mode labels.
            Transitions polarized along the $x$, $y$ and $z$ axes are shown by the markers
            $\bullet$, $\blacktriangle$ and $\blacksquare$, respectively.
            The appearance of the transitions $\textrm{GS} \rightarrow 2\nu_4$ and $\textrm{GS} \rightarrow 3\nu_4$
            demonstrate the anharmonicity of the potential along the $\nu_4$ mode.}
   \label{fig:spectrum_S1_ff_gs}
\end{figure}

A detailed analysis of the light-dressed spectrum in Fig. \ref{fig:spectrum_dressed1} has revealed that the peaks
below $1100 ~ \textrm{cm}^{-1}$ can be interpreted as transitions (both absorption and induced emission) from the 
vibrational eigenstates of A at $32931.2 ~ \textrm{cm}^{-1}$ and $32937.2 ~ \textrm{cm}^{-1}$
(both of $\textrm{B}_1$ symmetry) to other A vibrational eigenstates making up the final field-dressed states of
eqn \eqref{eq:final_state2}. The following qualitative conclusions have been drawn
by investigating one-dimensional wave function cuts of the relevant A vibrational eigenstates:
(a) the anharmonic nature of the $V_{\textrm{A}}$ PES often leads to multiple
changes in the vibrational quantum number $v_4$ ($|\Delta v_4| = 0,1,2,3,\dots$),
often accompanied by simultaneous changes in $v_2$, $v_3$ and $v_6$;
(b) the $\nu_1$ and $\nu_5$ modes hardly play any role;
(c) the $\textrm{g}_1$ and $\textrm{g}_2$ transitions ($x$ polarization) occur between A vibrational eigenstates
of $\textrm{B}_1$ and $\textrm{A}_1$ symmetries (the $x$ component of the PDM is of $\textrm{B}_1$ symmetry) and
$\Delta v_4$ is always odd due to symmetry;
(d) similarly, the $\textrm{g}_3$ ($y$ polarization) and $\textrm{g}_4$ ($z$ polarization) peaks involve
transitions of $\textrm{B}_1 \rightarrow \textrm{A}_2$ and $\textrm{B}_1 \rightarrow \textrm{B}_1$ types, respectively,
and $\Delta v_4$ is always even in both cases;
(e) the emergence of peaks in the interval $0-1100 ~ \textrm{cm}^{-1}$ can be understood as intensity borrowing
according to the model introduced in Section \ref{sec:lici}.
The intensity is borrowed from peaks that are also present in the field-free vibrational spectrum.
In light of these conclusions, the richness of the field-dressed spectrum is primarily attributed to
the $\nu_4$ mode. This statement is also corroborated by the field-free vibrational spectrum of H$_2$CO in its 
A electronic state, shown in Fig. \ref{fig:spectrum_S1_ff_gs}.
In Fig. \ref{fig:spectrum_S1_ff_gs} the initial state is chosen as the A vibrational
ground state (GS) and therefore the field-free vibrational spectrum in Fig. \ref{fig:spectrum_S1_ff_gs}
is not directly relevant for our current analysis.
It is apparent that, in contrast to the field-free vibrational spectrum of Fig. \ref{fig:spectrum_ff},
the anharmonicity of the $V_{\textrm{A}}$ PES along the $\nu_4$ mode allows multiple changes in $v_4$, as justified by the
appearance of the transitions $\textrm{GS} \rightarrow 2\nu_4$ and $\textrm{GS} \rightarrow 3\nu_4$
besides the fundamental transition $\textrm{GS} \rightarrow \nu_4$.

\begin{figure}[hbt!]
 \centering
   \includegraphics[scale=0.65]{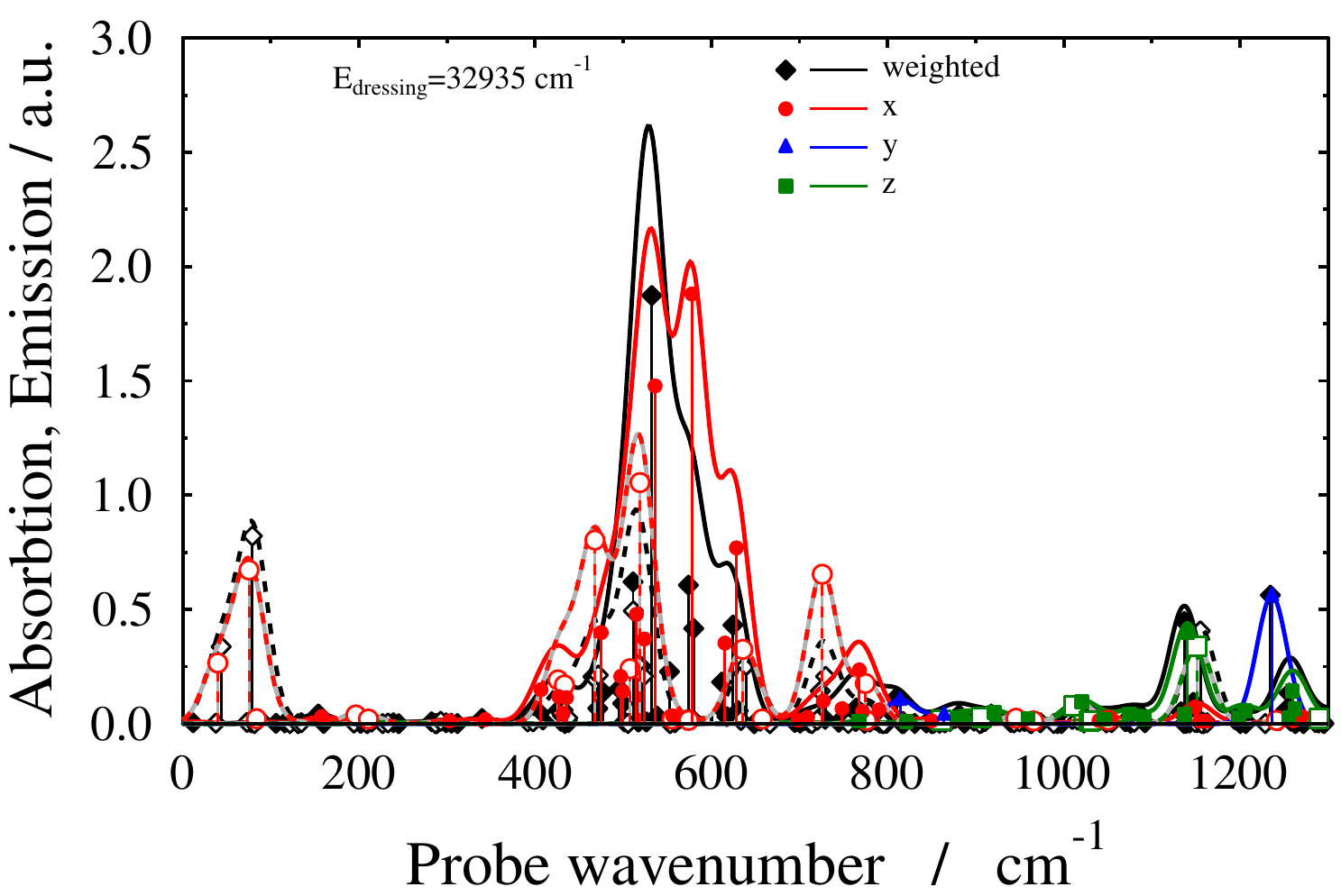}
   \includegraphics[scale=0.65]{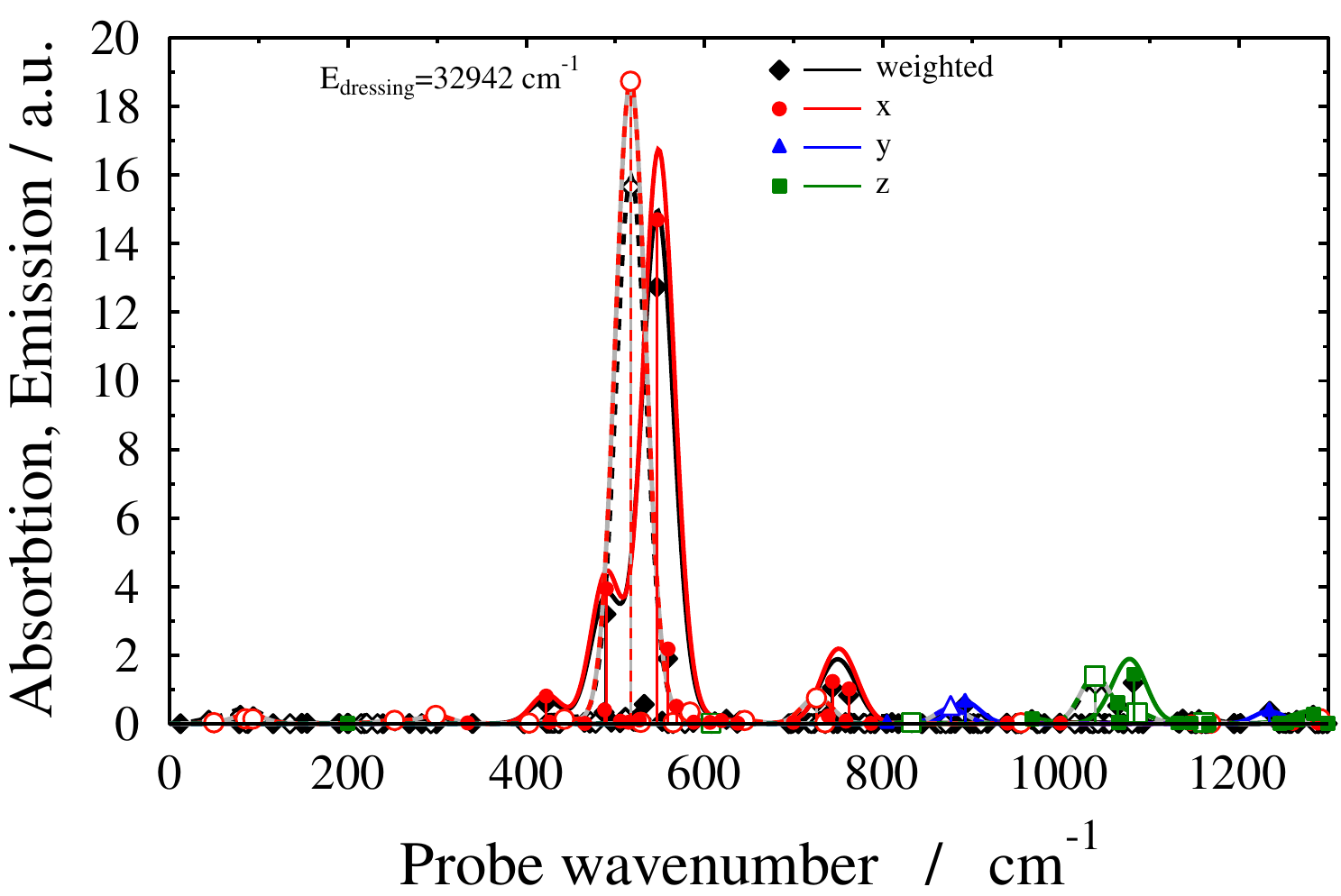}
   \caption{Absorption (solid line, full markers) and stimulated emission (dashed line, empty markers)
   	    spectra of H$_{2}$CO dressed with a laser field linearly polarized along the
            body-fixed $y$ axis with $I_\textrm{d} = 10^{11} ~ \textrm{W}/\textrm{cm}^2$,
            $\omega_\textrm{d} = 32935.0 ~ \textrm{cm}^{-1}$ (upper panel) and
            $\omega_\textrm{d} = 32942.0 ~ \textrm{cm}^{-1}$ (lower panel).
            Transitions polarized along the $x$, $y$ and $z$ axes are shown
            by the markers $\bullet$, $\blacktriangle$ and $\blacksquare$, respectively.
            The weighted field-dressed spectrum (see text) is indicated by the marker \ding{117}.
            Stick spectra were convolved with Gaussian functions having a full
            width at half maximum of $40 ~ \textrm{cm}^{-1}$.
            The most salient features of the field-dressed spectrum, as opposed to the field-free vibrational spectrum,
            are the appearance of numerous peaks below $1100 ~ \textrm{cm}^{-1}$
            and the sensitivity of the spectrum to the frequency of the dressing laser.}
   \label{fig:spectrum_dressed2}
\end{figure}

Two other interesting field-dressed spectra can be seen in Fig. \ref{fig:spectrum_dressed2}
where the molecule is dressed with a laser field linearly polarized along the
body-fixed $y$ axis with $I_\textrm{d} = 10^{11} ~ \textrm{W}/\textrm{cm}^2$
and $\omega_\textrm{d} = 32935.0 ~ \textrm{cm}^{-1}$ ($\lambda_\textrm{d} = 303.63 ~ \textrm{nm}$) or
$\omega_\textrm{d} = 32942.0 ~ \textrm{cm}^{-1}$ ($\lambda_\textrm{d} = 303.56 ~ \textrm{nm}$).
Similarly to the field-dressed spectrum in Fig. \ref{fig:spectrum_dressed1}, the 
$\textrm{g}_1$, $\textrm{g}_2$, $\textrm{g}_3$ and $\textrm{g}_4$ peaks all appear in the spectra
of Fig. \ref{fig:spectrum_dressed2}. In addition, another group of peaks is visible
in the upper panel of Fig. \ref{fig:spectrum_dressed2}, corresponding to $x$-polarized
transitions below $100.0 ~ \textrm{cm}^{-1}$, involving $|\Delta \nu_4|=1,3$ and no change
in the remaining vibrational modes.
The composition of the inital field-dressed states for the field-dressed spectra shown
in Fig. \ref{fig:spectrum_dressed2} helps identify the most relevant X and A vibrational eigenstates:
X vibrational ground state ($\textrm{A}_1$, $66.4\%$),
A vibrational eigenstate of $32931.2 ~ \textrm{cm}^{-1}$ ($\textrm{B}_1$, $2.4\%$)
and $32937.2 ~ \textrm{cm}^{-1}$ ($\textrm{B}_1$, $30.9\%$)
($\omega_\textrm{d} = 32935.0 ~ \textrm{cm}^{-1}$),
and
X vibrational ground state ($\textrm{A}_1$, $43.0\%$),
A vibrational eigenstate of $32937.2 ~ \textrm{cm}^{-1}$ ($\textrm{B}_1$, $9.3\%$)
and $32943.2 ~ \textrm{cm}^{-1}$ ($\textrm{B}_1$, $47.6\%$)
($\omega_\textrm{d} = 32942.0 ~ \textrm{cm}^{-1}$).
It is worth noting that Figs. \ref{fig:spectrum_dressed1} and \ref{fig:spectrum_dressed2}
demonstrate how sensitive the field-dressed spectrum can be to small changes
in $\omega_\textrm{d}$. As the dressing field with the current $\omega_\textrm{d}$ values couples
the X vibrational ground state to more close-lying A vibrational eigenstates
(as shown in Fig. \ref{fig:pes_1D}), small changes
in $\omega_\textrm{d}$, equivalent to slight shifts in the relative positions of the
relevant X and A vibrational eigenstates, have a remarkable impact on the structure of the
initial field-dressed state and thus on the intensity borrowing patterns in the field-dressed spectrum.

\begin{figure}[hbt!]
 \centering
   \includegraphics[scale=0.65]{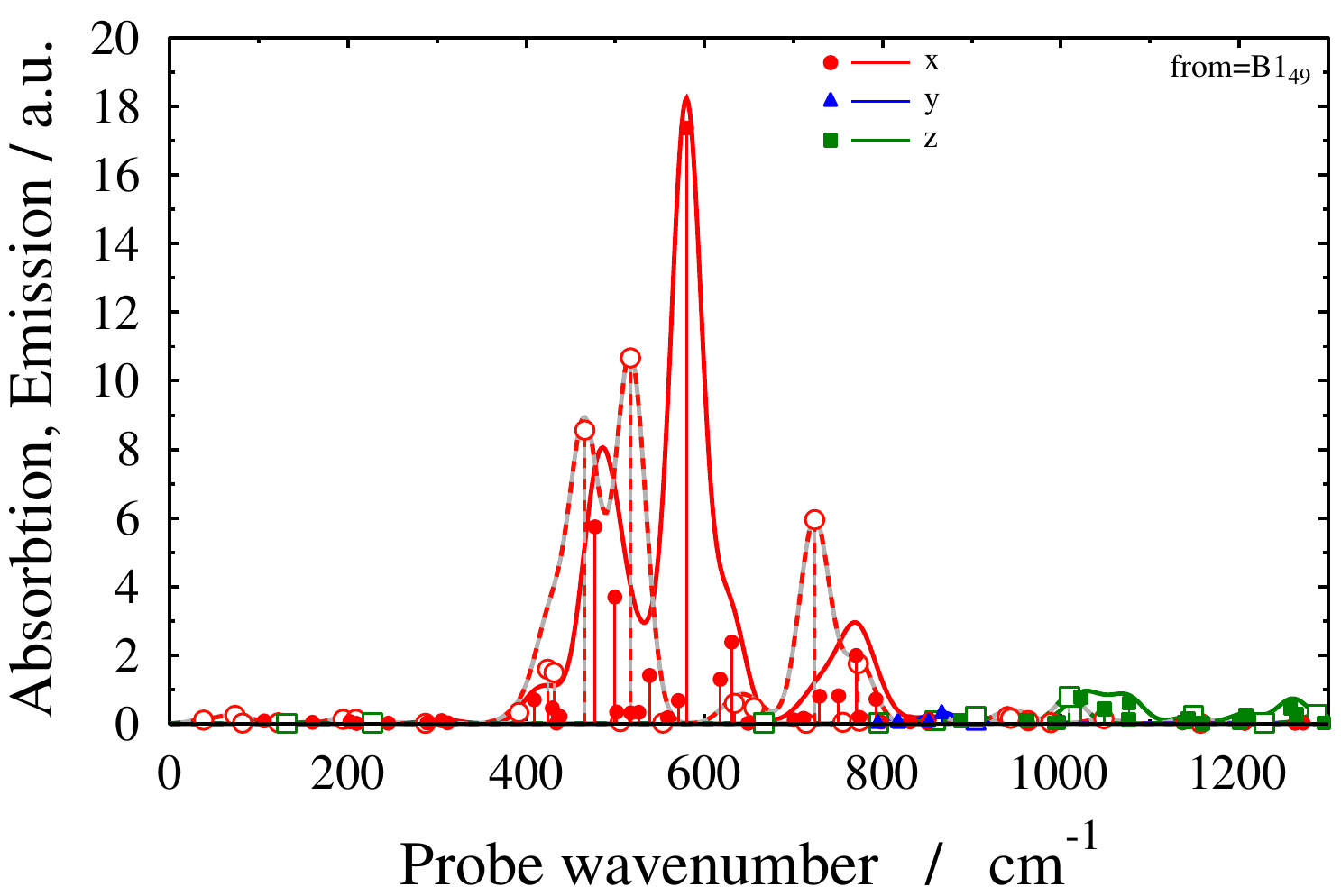}
   \includegraphics[scale=0.65]{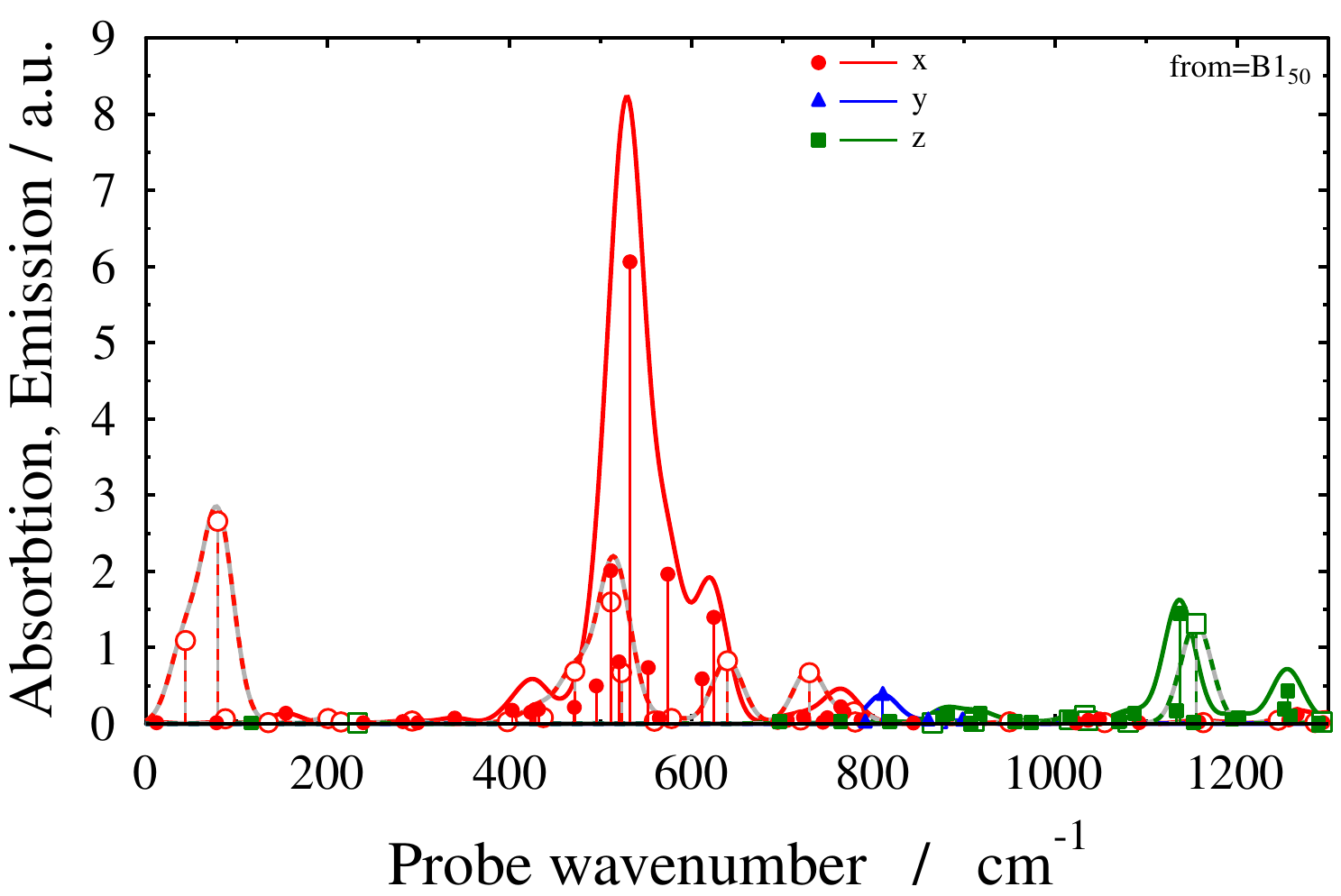}
   \includegraphics[scale=0.65]{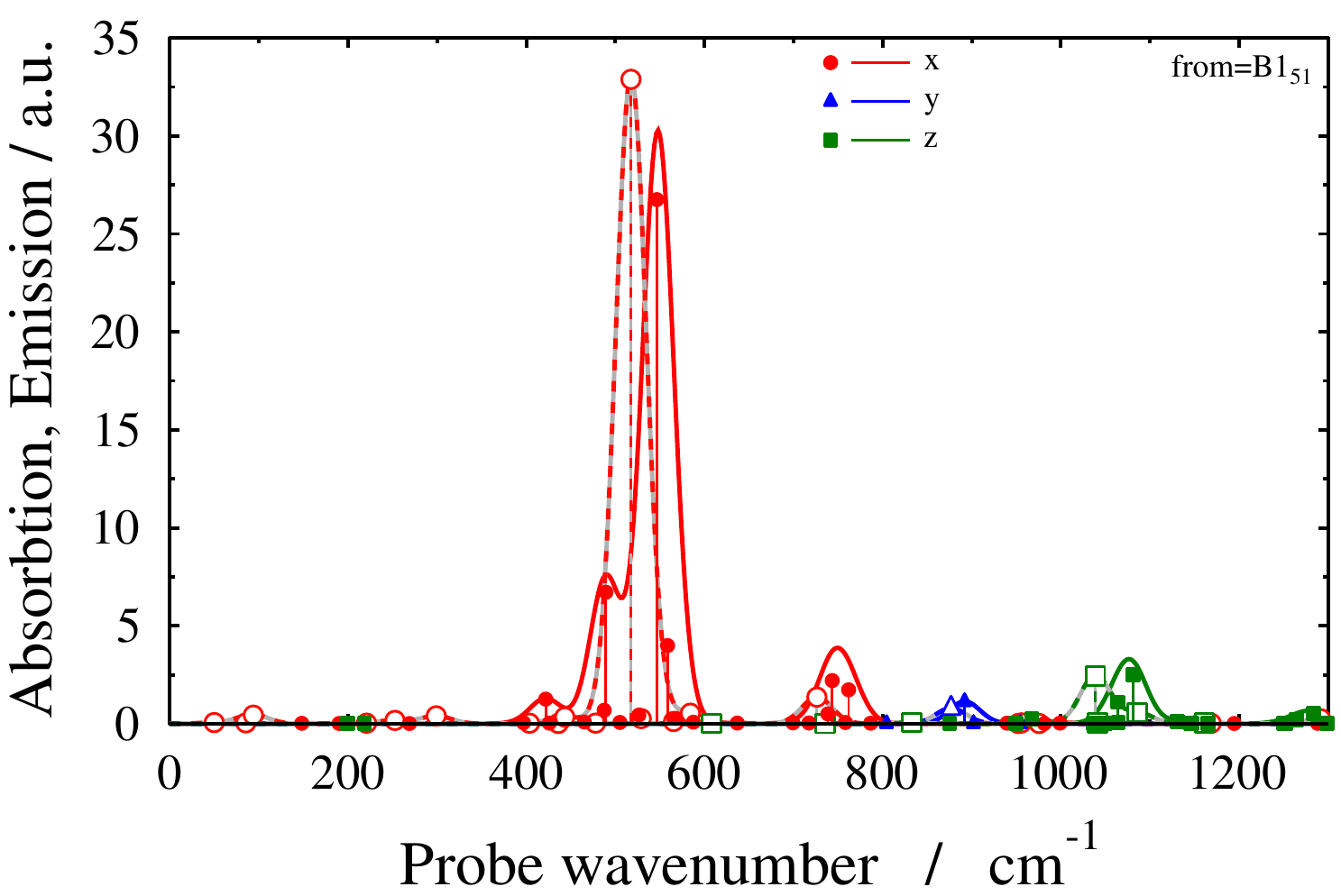}
   \caption{Field-free vibrational spectra (absorption: solid line with full markers,
   		stimulated emission: dashed line with empty markers)
   		of the H$_2$CO molecule in its excited electronic state (A).
            The different peaks correspond to transitions from an excited vibrational eigenstate
            ($32931.2 ~ \textrm{cm}^{-1}$: upper panel, $32937.2 ~ \textrm{cm}^{-1}$: middle panel,
             $32943.2 ~ \textrm{cm}^{-1}$: lower panel) to other vibrational eigenstates.
            Transitions polarized along the $x$, $y$ and $z$ axes are shown
             by the markers $\bullet$, $\blacktriangle$ and $\blacksquare$, respectively.
            Stick spectra were convolved with Gaussian functions having a full
            width at half maximum of $40 ~ \textrm{cm}^{-1}$.}
   \label{fig:spectrum_S1_ff_exc}
\end{figure}

The considerations made so far in this section are further supported by the
field-free vibrational spectra in Fig. \ref{fig:spectrum_S1_ff_exc} where the initial states have been
chosen as the A vibrational eigenstates of $32931.2 ~ \textrm{cm}^{-1}$, $32937.2 ~ \textrm{cm}^{-1}$
and $32943.2 ~ \textrm{cm}^{-1}$, having non-negligible populations in the previously
described inital field-dressed states.
As can be seen in Fig. \ref{fig:spectrum_S1_ff_exc}, several transitions, originating from
these eigenstates, appear below $1100 ~ \textrm{cm}^{-1}$ in the field-free spectra.
These transitions are strongly related to their field-dressed counterparts appearing in the
field-dressed spectra shown in Figs. \ref{fig:spectrum_dressed1} and \ref{fig:spectrum_dressed2}.

\begin{figure}[hbt!]
 \centering
   \includegraphics[scale=0.65]{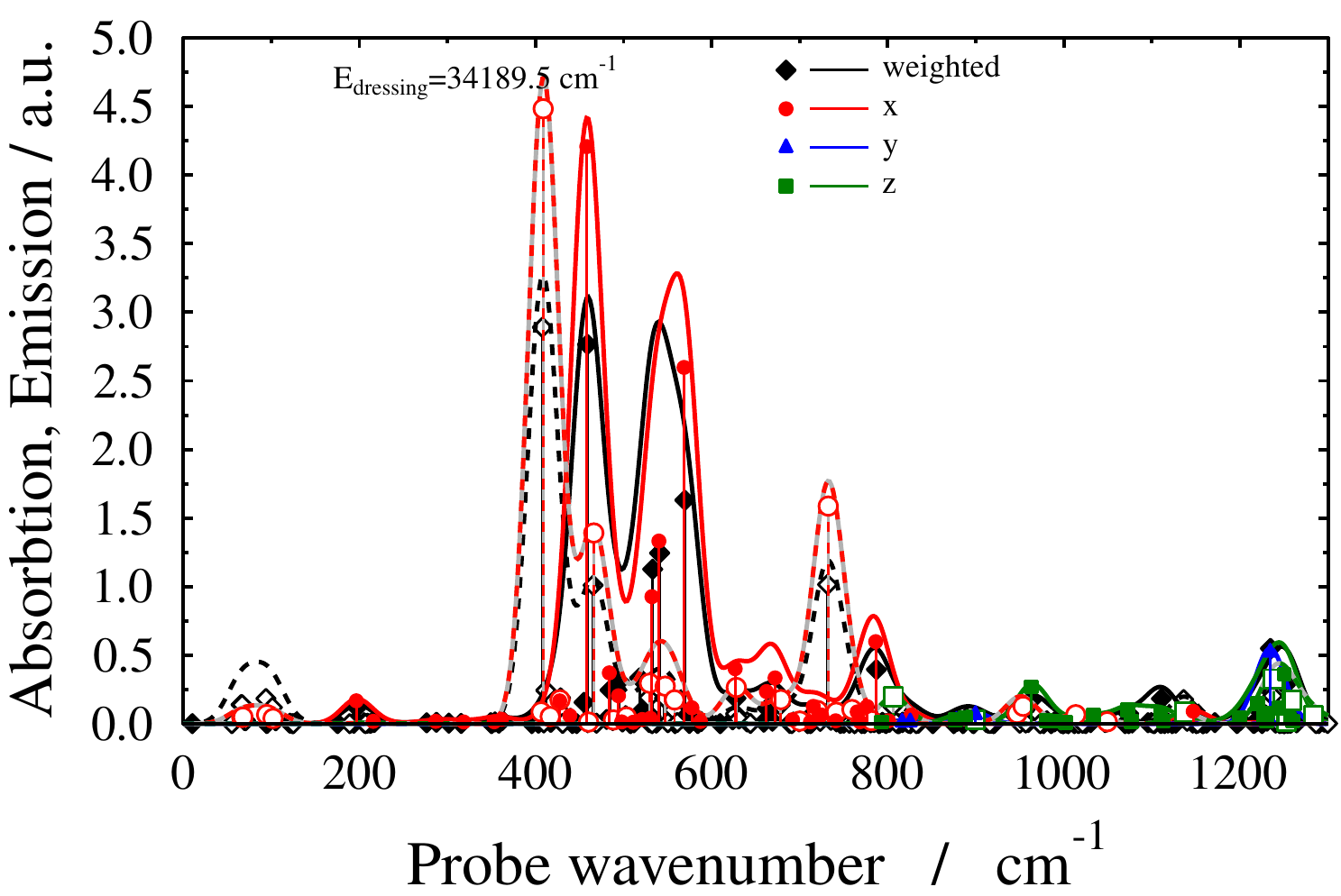}
   \includegraphics[scale=0.65]{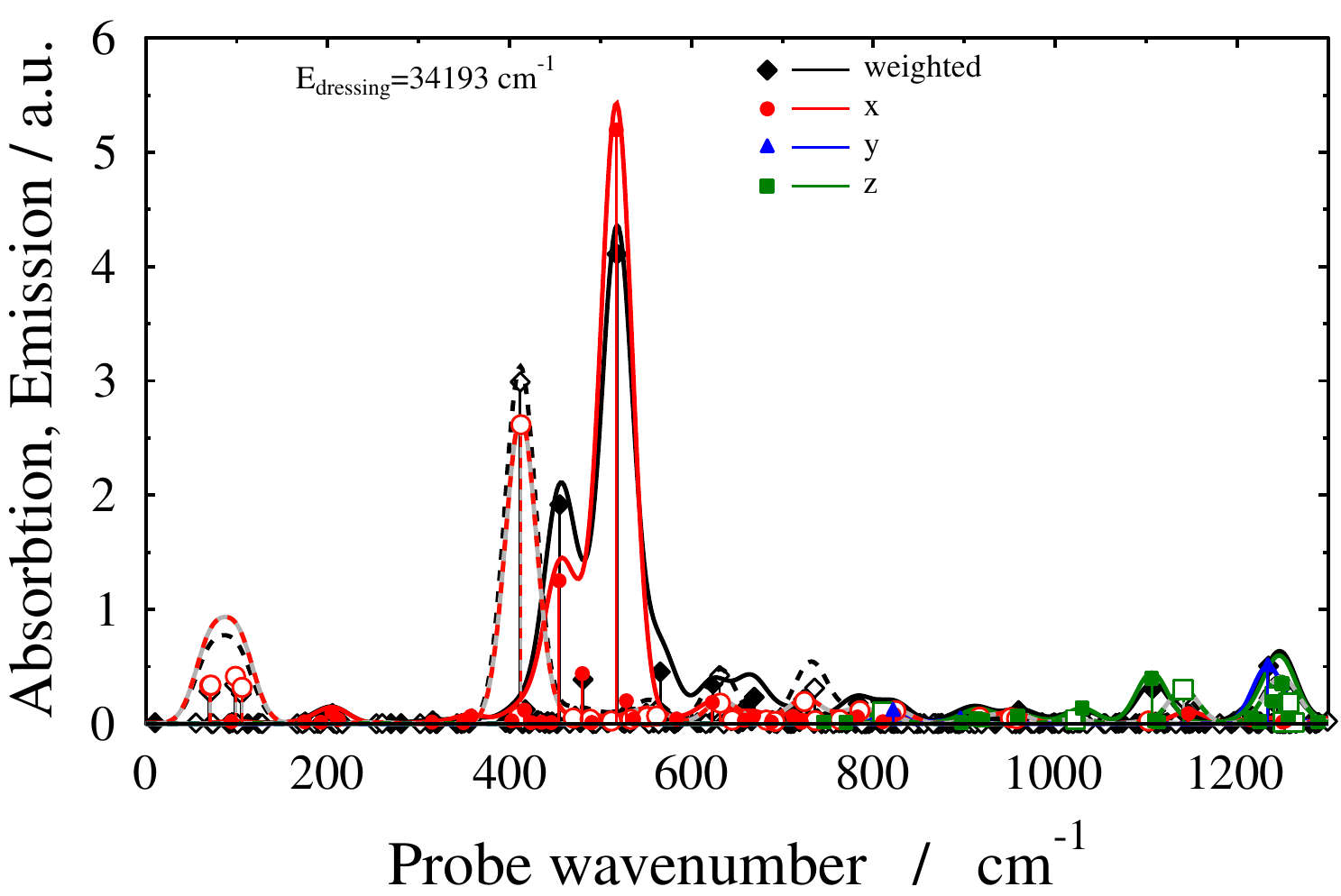}
   \includegraphics[scale=0.65]{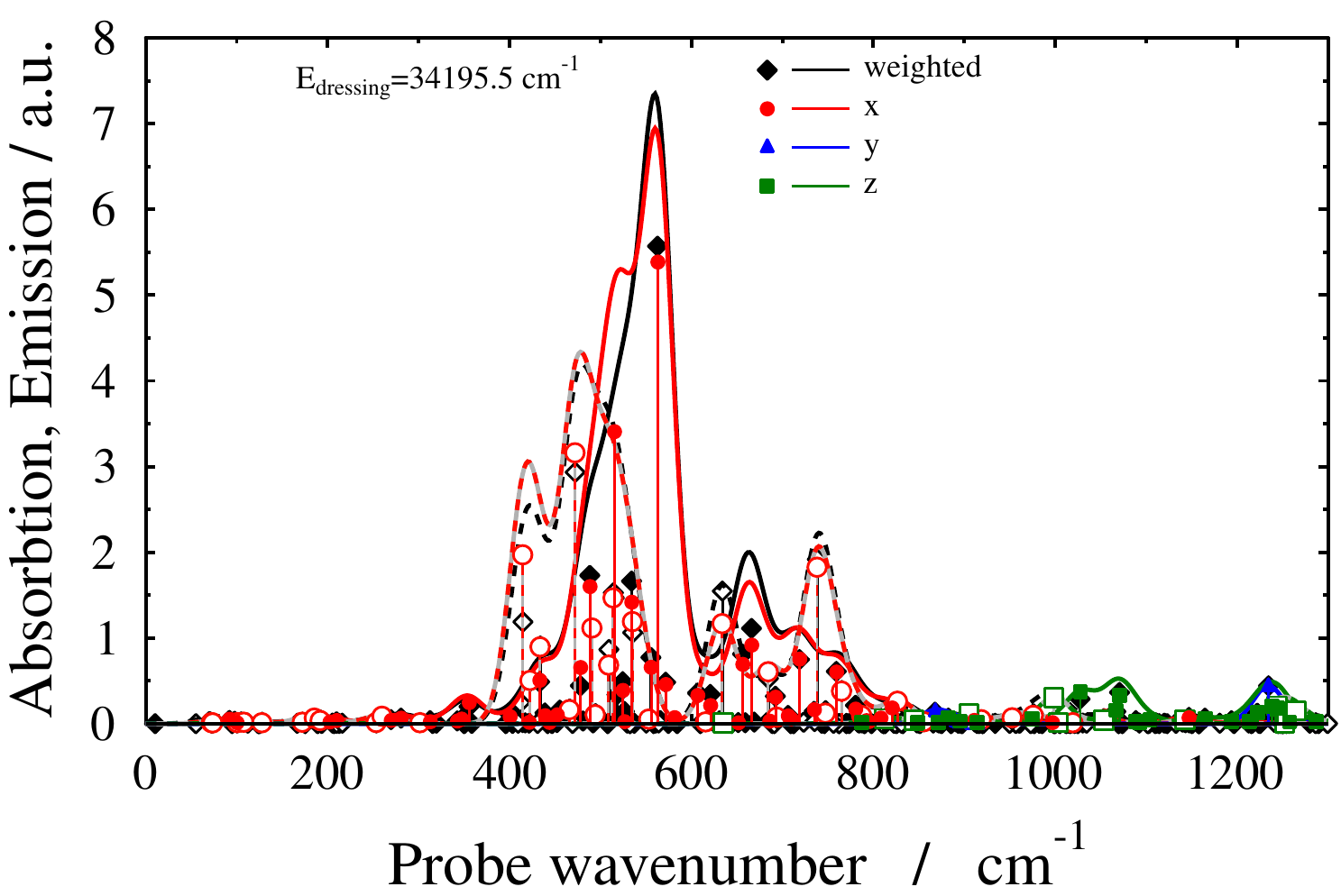}
   \caption{Absorption (solid line, full markers) and stimulated emission (dashed line, empty markers)
   		spectra of H$_{2}$CO dressed with a laser field linearly polarized along the
            body-fixed $y$ axis with $I_\textrm{d} = 10^{11} ~ \textrm{W}/\textrm{cm}^2$,
            $\omega_\textrm{d} = 34189.5 ~ \textrm{cm}^{-1}$ (upper panel),
            $\omega_\textrm{d} = 34193.0 ~ \textrm{cm}^{-1}$ (middle panel) and
            $\omega_\textrm{d} = 34195.5 ~ \textrm{cm}^{-1}$ (lower panel).
            Transitions polarized along the $x$, $y$ and $z$ axes are shown
            by the markers $\bullet$, $\blacktriangle$ and $\blacksquare$, respectively.
            The weighted field-dressed spectrum (see text) is indicated by the marker \ding{117}.
            Stick spectra were convolved with Gaussian functions having a full
            width at half maximum of $40 ~ \textrm{cm}^{-1}$.
            The most salient feature of the field-dressed spectrum, as opposed to the field-free vibrational spectrum,
            are the appearance of numerous peaks below $1100 ~ \textrm{cm}^{-1}$
            and the sensitivity of the spectrum to the frequency of the dressing laser.}
   \label{fig:spectrum_dressed3}
\end{figure}

Switching to larger dressing wavenumbers $\omega_\textrm{d}$, at certain values, similar
remarkable intensity borrowing patterns appear in the lower part of the field-dressed spectrum.
In Fig. \ref{fig:spectrum_dressed3}, the three field-dressed spectra with
dressing fields linearly polarized along the body-fixed $y$ axis with
$I_\textrm{d} = 10^{11} ~ \textrm{W}/\textrm{cm}^2$ and close-lying dressing wavenumbers of
$\omega_\textrm{d} = 34189.5 ~ \textrm{cm}^{-1}$ ($\lambda_\textrm{d} = 292.49 ~ \textrm{nm}$),
$\omega_\textrm{d} = 34193.0 ~ \textrm{cm}^{-1}$ ($\lambda_\textrm{d} = 292.46 ~ \textrm{nm}$) and
$\omega_\textrm{d} = 34195.5 ~ \textrm{cm}^{-1}$ ($\lambda_\textrm{d} = 292.44 ~ \textrm{nm}$) 
clearly show the prominent features of intensity borrowing, which can again be interpreted by
the considerations made earlier in this section.

Finally, one might notice that the dressing field has been polarized along
the body-fixed $y$ axis throughout our study, which was motivated by the strong
first-order $y$-oriented TDM produced upon displacement along the $\nu_4$ vibrational mode.
As the first derivatives of the $x$ component of the TDM with respect to the coordinates
are smaller than the corresponding values of the $y$ TDM component
and the first nonvanishing terms in the Taylor expansion of $z$ component of the TDM 
are bilinear, effects induced by dressing fields polarized along the body-fixed $y$ axis
are expected to be the most pronounced.
The investigation of effects caused by the orientation of the dressing polarization vector in
field-dressed spectra is left for future work.

\section{Conclusions}
In the present work we have outlined a pump-probe
scheme to obtain the field-dressed low-energy vibronic spectra of
polyatomic molecules treating all vibrational degrees of freedom.
The H$_2$CO molecule which does not exhibit a natural conical intersection
in the studied energy region was chosen as a showcase example.
In order to calculate the weak-field absorption and
stimulated emission spectra of the field-dressed H$_2$CO molecule we performed accurate,
six-dimensional computations for a multitude of dressing photon energies and dressing field intensities.

The results obtained clearly show the direct impact of the light-induced
conical intersection on the field-dressed spectra of the H$_2$CO molecule.
The findings are highly sensitive to variations of the dressing frequency which varies the location of the LICI,
shedding light on the mixing of the levels of different electronic states and on their coupling mechanism.
The emergence of peaks in the field-dressed spectrum
below $1100 ~ \textrm{cm}^{-1}$ undoubtedly demonstrates strong nonadiabatic
effects manifested in the so-called ``intensity borrowing'' phenomenon,
arising due to the presence of the light-induced nonadiabaticity.
The strong nonadiabaticity which is completely absent in
the field-free H$_2$CO molecule mixes the different electronic and
vibrational degrees of freedom creating a very rich pattern
in a certain interval of the low-energy vibronic spectrum
where no transitions occur in the field-free situation.
This intense mixing process could not happen without the presence of the
light-induced conical intersection.

\section*{Acknowledgements}
Professor Joel Bowman is gratefully acknowledged for providing Fortran subroutines for
the $\textrm{S}_0$ and $\textrm{S}_1$ potential energy surfaces.
This research was supported by the EU-funded Hungarian grant EFOP-3.6.2-16-2017-00005.
The authors are grateful to NKFIH for financial support (grants No. K128396 and PD124699).


\end{document}